\documentclass[epj]{svjour}

\usepackage[english]{babel}
\usepackage{blindtext}
\usepackage[utf8]{inputenc}
\usepackage{graphicx}
\usepackage{bm}
\usepackage{latexsym}
\usepackage{physics}
\usepackage{amsfonts}
\usepackage{bbold}
\usepackage{appendix}
\usepackage{xcolor}
\usepackage[normalem]{ulem}
\usepackage{feynmp-auto}

\usepackage[colorlinks = true,
            linkcolor = blue,
            urlcolor  = blue,
            citecolor = blue,
            anchorcolor = blue]{hyperref}

\definecolor{newgreen}{rgb}{0.2,0.6,0.1}

\usepackage{tikz}
\newcommand{\tikzm}[2]{
   \tikz[baseline=-0.65ex]{#2}
}
\tikzset{linePlain/.style={draw=black, thick}}

\newcommand{\eps}{\epsilon}
\newcommand{\up}{\uparrow}
\newcommand{\down}{\downarrow}
\newcommand{\bs}[1]{\mathbf{#1}}

\newcommand{\bk}{\mathbf{k}}
\newcommand{\bp}{\mathbf{p}}

\newcommand{\bQ}{\mathbf{Q}}

\newcommand\blfootnote[1]{%
  \begingroup
  \renewcommand\thefootnote{}\footnote{#1}%
  \addtocounter{footnote}{-1}%
  \endgroup
}

\usepackage{cite}
\title{Single-boson~exchange~functional renormalization group application to the two-dimensional Hubbard model at weak coupling}
\titlerunning{Single-boson~exchange~fRG application to the 2D Hubbard model}
\authorrunning{K. Fraboulet et al.}

\author{
	Kilian Fraboulet\inst{1,\dagger}
	\and
	Sarah Heinzelmann\inst{1,\dagger}
	\and
	Pietro M. Bonetti\inst{2}
	\and
	Aiman Al-Eryani\inst{1}
	\and
	Demetrio Vilardi\inst{2}
	\and
	Alessandro Toschi\inst{3}
	\and
	Sabine Andergassen\inst{1}
}

\institute{
	Institut f\"ur Theoretische Physik and Center for Quantum Science, Universit\"at T\"ubingen, Auf der Morgenstelle 14, 72076 T\"ubingen, Germany
	\and
	Max Planck Institute for Solid State Research, Heisenbergstrasse 1, D-70569 Stuttgart, Germany
	\and
	Institute of Solid State Physics, TU Wien, 1040 Vienna, Austria
}

\date{\today}

\abstract{
We illustrate the algorithmic advantages of the recently introduced single-boson exchange (SBE) formulation for the one-loop functional renormalization group (fRG), by applying it to the two-dimensional Hubbard model on a square lattice. We present a detailed analysis of the fermion-boson Yukawa couplings and of the corresponding physical susceptibilities by studying their evolution with temperature and interaction strength, both at half filling and finite doping. The comparison with the conventional fermionic fRG decomposition shows that the rest functions of the SBE algorithm, which describe correlation effects beyond the SBE processes, play a negligible role in the weak-coupling regime above the pseudo-critical temperature, in contrast to the rest functions of the conventional fRG. Remarkably, they remain finite also at the pseudo-critical transition, whereas the corresponding rest functions of the conventional fRG implementation diverge. As a result, the SBE formulation of the fRG flow allows for a substantial reduction of the numerical effort in the treatment of the two-particle vertex function, paving a promising route for future multiboson and multiloop extensions.
}

\begin{document}
\maketitle

\blfootnote{$^{\dagger}$These authors contributed equally to this work.}

\section{Introduction}

In the theoretical treatment of the two-particle interaction in strongly correlated electron systems, the recently introduced single-boson exchange (SBE) decomposition~\cite{Krien2019_I} represents a promising route for a computationally more efficient parametrization of the two-particle vertex~\cite{Krien2020,Krien2021}, whose numerical treatment often represents one of the main bottlenecks for advanced quantum many-body approaches. Differently to the well-known parquet decomposition~\cite{Dedominicis1964,Dedominicis1964A,Bickers2004,Yang2009,Tam2013,Rohringer2012,Valli2015,Kauch2019,Li2019}, the SBE classification is not based on the reducibility of two-particle vertex diagrams with respect to the cut of two fermionic propagators, but of the bare interaction $U$. As the electronic interaction is a two-particle operator, formal similarities can be found between the parquet and the SBE classifications of diagrams. For instance, a direct correspondence between the reducible classes (e.g., spin/charge particle-hole, and singlet/triplet particle-particle) diagrams of the two schemes can be easily identified. Therefore, similarly as in the parquet formalism, the diagrammatic structure of the SBE scheme includes the high-frequency asymptotic properties \cite{Rohringer2012,Tagliavini2018,Wentzell2016} of the corresponding irreducible diagrams. At the same time, the diagrammatic content of the respective reducible classes is quite different in the two descriptions. For this reason, the SBE classification of diagrams circumvents the problem of the multiple divergences of two-particle irreducible vertices that ubiquitously affects \cite{Schaefer2013,Janis2014,Gunnarsson2016,Schaefer2016,Ribic2016,Gunnarsson2017,Vucicevic2018,Chalupa2018,Springer2020,Chalupa2021} the theoretical description of many-electron models at intermediate-to-strong coupling, challenging, in particular, the applicability of parquet-based approaches in the nonperturbative regime.

For the functional renormalization group (fRG)\footnote{See Refs.~\cite{Metzner12} and \cite{Dupuis2021} for reviews.}, several bosonization procedures have been already proposed~\cite{Baier2004,Diehl2007,Diehl2007_II,Strack2008,Bartosch2009,Friederich2010,Friederich2011,Obert2013}. The application of the SBE formalism to the fRG put forward in Ref.~\cite{Bonetti2022} relies on the partial bosonization of the vertex function~\cite{Krahl2007,Friederich2010,Denz2020}, similarly to its use in the channel decomposition~\cite{Karrasch2008,Husemann2009,Wang2012,Vilardi2017,Tagliavini2019,Vilardi2019,Stepanov2019,Hille2020,Harkov2021} (as well as for parquet solvers~\cite{Eckhardt2020,Krien2020,Astretsov2020,Astleitner2020,Rohringer2020,Krien22}). In addition to the screened interaction, a fermion-boson Yukawa coupling~\cite{Schmalian1999,Sadovskii2019} (or Hedin vertex~\cite{Hedin1965}) is  determined from the vertex asymptotics, in analogy to the construction of the kernel functions describing the high-frequency asymptotics~\cite{Wentzell2016}. Through their relation, the flow equations of the screened interaction and Yukawa coupling in the SBE representation can be obtained in a straightforward way.

Here, we provide a systematic analysis of the two-dimensional (2D) Hubbard model\footnote{See Ref.~\cite{Qin21} for a recent overview of computational results for the 2D Hubbard model.} at weak coupling. In particular, we investigate the quality of the approximations that can arise from this decomposition in different parameter regimes, notably by comparing with the conventional fermionic formalism. An important result is that the divergent behavior arising in proximity of the pseudo-critical temperature, affects exclusively the screened interaction while the Yukawa couplings and the rest functions, which describe correlation effects beyond the SBE processes, remain finite. Differently than in the conventional fermionic fRG implementation, the SBE rest functions turn out to play a negligible role in the weak-coupling regime above the pseudo-critical transition. Beyond this remarkable numerical advantage, the SBE decomposition also provides a more natural description of the underlying physical content, allowing for a clear physical identification of the relevant degrees of freedom.

The paper is organized as follows. In Section~\ref{sec:method} we introduce the Hubbard model and briefly review the SBE decomposition together with its fRG implementation. We then present the results, at half filling in Section \ref{sec:results half filling} and at finite doping in Section \ref{sec:results doped regime}, for i) the most important quantities inherent to the SBE formalism, (i.e. the Yukawa couplings); ii) the physical observables (i.e. the susceptibilities); iii) the rest functions, which encompass the corrections beyond the SBE contributions. Analysing their relevance in the different parameter regimes we illustrate the advantages of the SBE-based fRG as compared to the conventional fermionic implementation of the fRG based on the parquet decomposition and high-frequency asymptotics. We finally conclude with a summary and an outlook in Section~\ref{sec:concl}.

\section{Model and method}
\label{sec:method}
%
\subsection{Hubbard model}

We consider the single-band Hubbard model in 2D,
\begin{equation}
    \begin{split}
        \mathcal{H}=\sum_{i\neq j,\sigma}t_{ij}c^\dagger_{i\sigma}c_{j\sigma}+U\sum_i n_{i\up}n_{i\down}-\mu\sum_{i,\sigma} n_{i\sigma},
    \end{split}
    \label{eq: Hubbard model}
\end{equation}
where $c_{i\sigma}$ ($c^{\dagger}_{i\sigma}$) annihilates (creates) an electron with spin $\sigma$ at the lattice site $i$ ($n_{i\sigma}=c^{\dagger}_{i\sigma}c_{i\sigma}$), $t_{ij}=-t$ is the hopping between nearest-neighbor sites, $t_{ij}=-t'$ the hopping between next-nearest-neighbor sites, $U$ the on-site Coulomb interaction, and $\mu$ the chemical potential.

In the following we use $t\equiv 1$ as the energy unit.
\subsection{Single-boson exchange decomposition}

We here briefly review the SBE decomposition and its application to the fRG~\cite{Bonetti2022},  providing the (one-loop) flow equations.

In a translationally invariant system with  U(1)-charge and SU(2)-spin symmetries, the two-particle vertex can be expressed as~\cite{Rohringer2012}
\begin{equation}
    \begin{split}
        V_{\sigma_1\sigma_2\sigma_3\sigma_4}(k_1,k_2,k_3,k_4) & =
        V(k_1,k_2,k_3)\delta_{\sigma_1\sigma_3}\delta_{\sigma_2\sigma_4}\\
        &-V(k_2,k_1,k_3)\delta_{\sigma_1\sigma_4}\delta_{\sigma_2\sigma_3},
    \end{split}
    \label{eq: SU(2) U(1) vertex}
\end{equation}
with $k_4=k_1+k_2-k_3$ due to energy and momentum conservation. $\sigma_i$ represents the spin quantum number and $k_i=(\bk_i,\nu_i)$ includes both the momentum and Matsubara frequency. According to the SBE decomposition, Eq.~\eqref{eq: SU(2) U(1) vertex} can be expressed through
\begin{equation}
    \begin{split}
        V(k_1,k_2,k_3) & =
        \mathcal{I}_{U_\mathrm{irr}}(k_1,k_2,k_3) + \nabla^\mathrm{M}_{k_1 k_3}(k_2-k_3)\\
                        & +\frac{1}{2}\left[ \nabla^\mathrm{M}_{k_1 k_4}(k_3-k_1)+\nabla^\mathrm{D}_{k_1 k_4}(k_3-k_1)\right]\\
                       &  + \nabla^{\mathrm{SC}}_{k_1 k_3}(k_1+k_2)- 2U,
    \end{split}
    \label{eq: vertex parametrization}
\end{equation}
where $\mathcal{I}_{U_\mathrm{irr}}$ accounts for the fully $U$-irreducible diagrams with respect to the removal of a bare interaction vertex that cuts the diagram into two disconnected parts. Among the $U$-reducible diagrams, we can identify three different channels, depending on how the fermionic legs are connected to the removed interaction $U$: $\nabla^\mathrm{M}$ includes all $U$-reducible diagrams in the particle-hole-crossed, $(\nabla^\mathrm{M}+\nabla^\mathrm{D})/2$ in the particle-hole, and $\nabla^{\mathrm{SC}}$ in the particle-particle channel (the last contribution of $-2U$ in Eq.~\eqref{eq: vertex parametrization} compensates for the double counting of the bare interaction already included in the $U$-reducible channels). The functions $\nabla^\mathrm{X}$, with $\mathrm{X}=\mathrm{M}$, $\mathrm{D}$ and $\mathrm{SC}$ corresponding to the magnetic, density and superconducting channels respectively, depend on two fermionic variables (in the indices) and a bosonic one (in the brackets). These effective interactions between two fermions mediated by the exchange of a boson can be expressed by 
\begin{equation}
    \label{eq: bosonic decomposition}
        \nabla^\mathrm{X}_{kk'}(Q)\equiv \lambda^\mathrm{X}_{k}(Q)\,w^\mathrm{X}(Q)\,\Bar{\lambda}^\mathrm{X}_{k'}(Q),
\end{equation}
where their dependence on the fermionic arguments has been factorized. The bosonic arguments are indicated by capital letters in the following, with $Q=(\bQ,\Omega)$. The screened interaction $w$, which plays the role of an effective bosonic propagator, is related to the physical susceptibility~\cite{Krien2019_I,Krien2019_II} by
\begin{equation}
\label{eq: D}
         w^\mathrm{X}(Q)= U -(\mathrm{sgn}\mathrm{X}) U^2 \chi^\mathrm{X}(Q), 
\end{equation}
where $\mathrm{sgn}\,\mathrm{M}=-1$, and $\mathrm{sgn}\,\mathrm{D}=\mathrm{sgn}\,\mathrm{SC}=1$ for the magnetic, density, and superconducting susceptibilities respectively. $\lambda$ ($\Bar{\lambda}$) is the left(right)-sided Yukawa coupling. Using the symmetries under the simultaneous exchange of the ingoing and outgoing variables, as well as under the simultaneous exchange of the two ingoing variables with the two outcoming ones, one can show that $\lambda=\Bar{\lambda}$ for the convention introduced in Eq.~\eqref{eq: vertex parametrization}.

Finally, the corresponding rest functions $M^\mathrm{X}$ containing the multiboson exchange contributions (which can not be represented in terms of $\lambda$ and $w$ alone) are identified by the $U$-irreducible contribution to the two-particle reducible channel
\begin{equation}
    \begin{split}
        \mathcal{I}_{U_\mathrm{irr}}(k_1,k_2,k_3) & =
        \mathcal{I}(k_1,k_2,k_3) - U + M^\mathrm{M}_{k_1 k_3}(k_2-k_3)\\
                        & +\frac{1}{2}\left[ M^\mathrm{M}_{k_1 k_4}(k_3-k_1)+M^\mathrm{D}_{k_1 k_4}(k_3-k_1)\right]\\
                       &  + M^{\mathrm{SC}}_{k_1 k_3}(k_1+k_2),
    \end{split}
\end{equation}
where $\mathcal{I}$ is the fully two-particle irreducible vertex. Note that for transfer momentum $\bs{Q} = \bs{0}$, all two-particle reducible diagrams in the particle-particle channel that exhibit a $d$-wave symmetry in the secondary momentum dependence on $k$ and $k'$ are in fact $U$-irreducible (see also Section \ref{sec: fermionic formalism}). As a consequence, at $\bs{Q}=\bs{0}$ the $d$-wave superconducting channel consists only of the rest function $M^{\mathrm{SC},d}$. 

The asymptotic high-frequency behavior is characterized by
\begin{subequations}
    \begin{align}
    \lim_{\Omega\rightarrow\infty} w^\mathrm{X}(\bs{Q},\Omega) &= U,\\
    \lim_{\Omega\rightarrow\infty} \lambda^\mathrm{X}_{k}(\bs{Q},\Omega) &= 1,\\
    \lim_{\Omega\rightarrow\infty} M^\mathrm{X}_{kk'}(\bs{Q},\Omega)& = 0,
    \end{align}
\end{subequations}    
for the bosonic frequency, whereas for large fermionic frequencies holds 
\begin{subequations}
    \begin{align}
    \lim_{\nu\rightarrow\infty} \lambda^\mathrm{X}_{(\bk,\nu)}(Q)& = 1,\\
    \lim_{\nu\rightarrow\infty} M^\mathrm{X}_{(\bk,\nu),k'}(Q)& = \lim_{\nu'\rightarrow\infty}  M^\mathrm{X}_{k,(\bk',\nu')}(Q)=0.
    \end{align}
\end{subequations}

\subsection{Single-boson exchange fRG}
\label{sec:SBE}

The fRG implementation \cite{Metzner12} relies on the truncation of the infinite hierarchy of flow equations for the $n$-particle irreducible vertex functions at the two-particle level. Neglecting the renormalization of three- and higher-order particle vertices yields approximate one-loop flow equations for the self-energy and two-particle vertex\footnote{More strongly correlated parameter regimes become accessible by exploiting the dynamical mean-field theory as starting point for the fRG flow \cite{Taranto2014,Vilardi2019}.}. The underlying approximations are devised for the weak to moderate coupling regimes.

In the SBE decomposition, the one-loop flow equations~\cite{Bonetti2022} for the screened interactions, Yukawa couplings, and rest functions, which we report here for completeness, read
\begin{subequations}
    \label{eq: flow eq general}
    \begin{align}
        \partial_\Lambda w^\mathrm{X}(Q)& = \left[w^\mathrm{X}(Q)\right]^2\!\!\int_p \lambda^\mathrm{X}_p(Q) \left[\widetilde{\partial}_\Lambda \Pi^\mathrm{X}_p(Q) \right] \lambda^\mathrm{X}_p(Q), \label{eq: flow eq D general}\\
        \partial_\Lambda \lambda^\mathrm{X}_k(Q)& = \int_p \mathcal{I}^\mathrm{X}_{kp}(Q) \left[\widetilde{\partial}_\Lambda \Pi^\mathrm{X}_p(Q) \right] \lambda^\mathrm{X}_p(Q), \label{eq: flow eq h general}\\
        \partial_\Lambda M^\mathrm{X}_{kk'}(Q) &= \int_p \mathcal{I}^\mathrm{X}_{kp}(Q) \left[\widetilde{\partial}_\Lambda \Pi^\mathrm{X}_p(Q) \right] \mathcal{I}^\mathrm{X}_{pk'}(Q), \label{eq: flow eq R general}
    \end{align}
\end{subequations}
where the explicit dependence on the RG scale $\Lambda$ of the various functions is omitted for simplicity. The symbol $\int_p=T\sum_\nu\int_\mathrm{B.Z.}\frac{d^2\bp}{(2\pi)^2}$ denotes the sum over fermionic Matsubara frequencies and a momentum integration over the Brillouin zone. The bubbles are defined by 
\begin{subequations}
    \begin{align}
        \Pi^\mathrm{M}_k(Q)& = -G(k)G(k+Q),\\
        \Pi^\mathrm{D}_k(Q)& = G(k)G(k+Q),\\
        \Pi^{\mathrm{SC}}_k(Q)& = -G(k)G(Q-k),
    \end{align}
\end{subequations}
where the inverse propagator including the self-energy $\Sigma$ is determined by the Dyson equation
\begin{equation}
    G^{-1}(k)=\left(\frac{\Theta^\Lambda(k)}{i\nu-\eps_\bk+\mu}\right)^{-1}-\Sigma(k).
\end{equation}
The symbol $\widetilde{\partial}_\Lambda$ denotes the derivative with respect to the explicit RG scale dependence of the propagator introduced by the cutoff function $\Theta^\Lambda(k)$. The functions 
\begin{equation}
    \mathcal{I}^\mathrm{X}_{kk'}(Q)=I^\mathrm{X}_{kk'}(Q)-\lambda^\mathrm{X}_{k}(Q)\,w^\mathrm{X}(Q)\,\lambda^\mathrm{X}_{k'}(Q),
\end{equation}
with 
\begin{subequations}
    \begin{align}
        I^\mathrm{M}_{kk'}(Q)& = V(k',k,k+Q),\label{eq: Lm}\\
        I^\mathrm{D}_{kk'}(Q)& = 2V(k,k',k+Q) - V(k',k,k+Q),\label{eq: Lc} \\
        I^{\mathrm{SC}}_{kk'}(Q)& = V(k,Q-k,k'),\label{eq: Ls}
    \end{align}
\end{subequations}
on the right-hand side of Eqs.~\eqref{eq: flow eq general} determine the flow of the Yukawa couplings as well as of the rest functions. The initial conditions at $\Lambda=\Lambda_\text{ini}$ read 
\begin{subequations}
    \label{eq: init frg}
    \begin{align}
        w^\mathrm{X}_\text{ini}(Q)& = U, \\
        \lambda^\mathrm{X}_{\text{ini},k}(Q)& = 1,\\
        M^\mathrm{X}_{\text{ini},kk'}(Q)& = 0,
    \end{align}
\end{subequations}
which, by comparing with Eq.~\eqref{eq: vertex parametrization}, is equivalent to imposing $V_{\text{ini}}=U$.

Neglecting the rest function, as explored in the results section, amounts to neglecting its flow in Eqs.~\eqref{eq: flow eq general}. This is not to be confused with the approximation put forward in Ref.~\cite{Gievers22}, where the rest functions are set to zero prior to derivation of the flow equations. The latter leads to simpler flow equations omitting entire classes of diagrams for the screened interactions and Yukawa couplings. The differences between these related approximations are detailed in the Appendix~\ref{sec: app2}.

The above Eqs.~\eqref{eq: flow eq general} are supplemented by the self-energy flow 
\begin{equation}
    \partial_\Lambda\Sigma(k) = \int_{p} \left[ 2V(k,k,p) - V(p,k,k) \right]S(p),
    \label{eq: sig flow}
\end{equation}
with the single-scale propagator $S=\partial_\Lambda G|_{\Sigma=\text{const}}$ and the initial condition $\Sigma_\text{ini}=0$.

We here use a frequency cutoff \cite{Husemann2009,Husemann2012} in the bare propagator $G_0^\Lambda(\bk,\nu) = \Theta^\Lambda(\nu)G_0(\bk,\nu) $, with $\Theta^\Lambda(\nu) =\nu^2/(\nu^2+\Lambda^2)$. For the parametrization of the two-particle vertex, we combine the truncated-unity fRG~\cite{Husemann2009,Wang2012,Lichtenstein2017} using the channel decomposition in conjunction with a form factor expansion for the momentum dependence with the full frequency treatment~\cite{Vilardi2017}, which includes the high-frequency asymptotics~\cite{Wentzell2016,Rohringer2012}. The flow equations in the form factor expansion are given in the Appendix~\ref{sec: flow eq in form factors}. For the details of the algorithmic implementation we refer to Refs.~\cite{Tagliavini2019,Hille2020}, the technical parameters are reported in Table~\ref{tab:TechnicalParameters}.

\begin{table}[t]
    \centering
    \begin{tabular}{|ccc|ccc|ccc|ccc|}
         \hline
         & $T$ & & & $n$ & & & $k_x$ & & &  $p_x$ & \\
         \hline
         & $0.1$ & & & $8$ & & & $18$  & & & $90$ & \\
         \hline 
         & $0.15$ & & & $8$ & & & $16$  & & & $80$ & \\
         \hline
         & $0.2 - 0.4$ & & & $6$ & & & $16$  & & & $80$ & \\
         \hline         
    \end{tabular}
    \caption{Specific parameters used for the technical implementation of our fRG calculations. The integer $n$ is the number of positive fermionic frequencies for which the two-particle vertex is evaluated. The screened interactions $w^{\mathrm{X}}$ are computed for $128n+1$ frequencies, whereas the Yukawa couplings $\lambda^{\mathrm{X}}$ for $(4n+1) \times 2n$ ones. The SBE rest functions are determined for $4n+1$ bosonic and $(2n)^2$ fermionic frequencies, with a total of $(4n+1) \times (2n)^2$. Furthermore, the fermionic frequency dependence of the self-energy is accounted for by $8n$ frequencies. For the dependence on the bosonic momentum $\bs{Q}$, the grid for the screened interactions, Yukawa couplings, and rest functions contains $(k_x\times k_x)$ momenta in the Brillouin zone, with a refinement of $N_k^\text{refine}=24$ additional points around $\bs{Q}=(\pi,\pi)$ to resolve the antiferromagnetic peak. The fermionic momentum dependence is accounted for by a form factor expansion, where we consider the (local) $s$-wave and at finite doping additionally the $d$-wave contribution. The momentum grid of the self-energies also spans over $(k_x\times k_x)$ momenta. Finally internal Green's function sums in the bubble and flow of the self-energy are performed on the finer grid with ($p_x\times p_x)$ momenta.}
    \label{tab:TechnicalParameters}
\end{table}

\subsection{Relation to the conventional fermionic formalism}
\label{sec: fermionic formalism}

As shown in Ref.~\cite{Bonetti2022}, the SBE formalism is related to the channel asymptotics~\cite{Wentzell2016} based on the parquet decomposition since they both rely on a similar classification of the diagrams contributing to the two-particle reducible vertex functions. 

For the screened interactions we obtain
\begin{equation}
    w^\mathrm{X}(Q)=U + \mathcal{K}^{(1)\mathrm{X}}(Q).
    \label{eq: screened interaction vs chi}
\end{equation}
The function $\mathcal{K}^{(1)\mathrm{X}}$ is defined by~\cite{Wentzell2016}
\begin{equation}
    \mathcal{K}^{(1)\mathrm{X}}(Q)=\lim_{\nu,\nu'\to\infty} \phi^\mathrm{X}_{(\bk,\nu),(\bk',\nu')}(Q),
\end{equation}
where $\phi^\mathrm{X}$ is the sum of all two-particle reducible diagrams in the $\mathrm{M}$, $\mathrm{D}$ or $\mathrm{SC}$ channel.

Similarly, the Yukawa coupling is related to $\mathcal{K}^{(2)\mathrm{X}}$ by
\begin{equation}
    \lambda^\mathrm{X}_k(Q) = 1 + \frac{\mathcal{K}^{(2)\mathrm{X}}_k(Q)}{w^\mathrm{X}(Q)},
\end{equation}
with 
\begin{equation}
    \mathcal{K}^{(2)\mathrm{X}}_k(Q)=\lim_{\nu'\to\infty} \phi^\mathrm{X}_{k,(\bk',\nu')}(Q) - \mathcal{K}^{(1)\mathrm{X}}(Q).
\end{equation}

Up to here, the SBE decomposition seems to offer no substantial computational gain as compared to the channel asymptotics, except for the situation where $\mathcal{K}^{(1)\mathrm{X}}$ and $\mathcal{K}^{(2)\mathrm{X}}$ acquire large values, i.e. in the vicinity of a pseudo-critical transition\footnote{Note that since the Mermin-Wagner theorem is not fulfilled in the one-loop approximation~\cite{Kugler2018_I,Tagliavini2019}, an AF transition is observed at a finite value of the interaction or temperature.}. In the SBE scheme, this occurs only for $w^\mathrm{X}$, while the Yukawa coupling always remains finite at weak coupling. However, the most important difference is visible in the rest functions: the SBE and the asymptotic rest functions are related by~\cite{Bonetti20}
\begin{equation}
    M^{\mathrm{X}}_{kk'}(Q)=\mathcal{R}^{\mathrm{X}}_{kk'}(Q)-[\lambda^\mathrm{X}_k(Q)-1]w^\mathrm{X}(Q)[\lambda^\mathrm{X}_{k'}(Q)-1],
    \label{eq: R_SBE vs R_asym}
\end{equation}
with 
\begin{equation}
    \mathcal{R}^{\mathrm{X}}_{kk'}(Q) = \phi^\mathrm{X}_{kk'}(Q) - \mathcal{K}^{(1)\mathrm{X}}(Q) - \mathcal{K}^{(2)\mathrm{X}}_{k}(Q) -\mathcal{K}^{(2)\mathrm{X}}_{k'}(Q),
\end{equation}
or, equivalently,
\begin{equation}
    M^{\mathrm{X}}_{kk'}(Q)=\phi^\mathrm{X}_{kk'}(Q) - \lambda^\mathrm{X}_{k}(Q)\,w^\mathrm{X}(Q)\,\lambda^\mathrm{X}_{k'}(Q)+U.
\end{equation}
Its inversion illustrates the relation between the parquet decomposition characteristic for the fermionic formulation and the SBE framework:
\begin{equation}
\phi^\mathrm{X}_{kk'}(Q) = \nabla^{\mathrm{X}}_{kk'}(Q)+ M^{\mathrm{X}}_{kk'}(Q) -U.
\end{equation}
In particular, the $d$-wave superconducting contribution to the effective interaction $\nabla^{\mathrm{SC},d}_{kk'}(Q)$ vanishes for transfer momentum $\bs{Q}=\bs{0}$~\cite{Heinzelmann22}. This implies that for the $d$-wave superconducting channel $M^{\mathrm{SC}}_{kk'}(\bs{0},\Omega)=\mathcal{R}^{\mathrm{SC}}_{kk'}(\bs{0},\Omega)$.

We finally note that, when including the flow of the rest functions, the SBE implementation is equivalent to the one based on the conventional fermionic fRG. The advantage lies in the substantial reduction of the numerical effort provided by the SBE formulation, as discussed below.

%
%
%
\section{Results at half filling}
\label{sec:results half filling}

To illustrate the potential of the SBE implementation for the fRG, we first consider the half-filled Hubbard model for $t'=0$, where the physical behavior is dominated by antiferromagnetic (AF) fluctuations. For this reason we include only the $s$-wave form factor to study the momentum and frequency as well as the temperature and interaction strength dependence of the Yukawa couplings and the susceptibilities in the different channels (in presence of $d$-wave form factors considered at finite doping, the corresponding components will be specified by an additional superscript). In particular, we will assess the importance of the rest functions for the accurate computation of physical observables (i.e. the susceptibilities and self-energies) and compare to data obtained by using the conventional fermionic formulation, see also Section~\ref{sec:SBE}. In order to make the comparison between the SBE and the conventional implementation more transparent, the parameter sets used in both cases include the same number of frequencies and momenta for each of the corresponding quantities ($\mathcal{K}^{(1)\mathrm{X}}$ and the screened interactions $w^{\mathrm{X}}$, and $\mathcal{K}^{(2)\mathrm{X}}$ and the Yukawa couplings $\lambda^{\mathrm{X}}$) which has been considered.

\subsection{Yukawa couplings}

We start by considering the quantities inherent to the SBE formulation. Since the screened interactions are directly related to the physical susceptibilities by Eq.~\eqref{eq: D} that will be discussed in the next section, we here consider the Yukawa couplings $\lambda^\mathrm{X}$. In this study, we analyse the accuracy of the SBE representation of the effective interaction when neglecting the flow of the rest functions ($M^\mathrm{X}=0$). Since the screened interactions and Yukawa couplings depend on less arguments than the full vertex, this allows for a significant gain both in memory cost as well as computational time with respect to the conventional fermionic formalism. By comparing the results obtained with and without the inclusion of the rest function, we investigate to what extent the approximation of neglecting its flow is justified in different parameter regimes.

\begin{figure}[t]
    \centering
    \includegraphics[width = 0.48\textwidth]{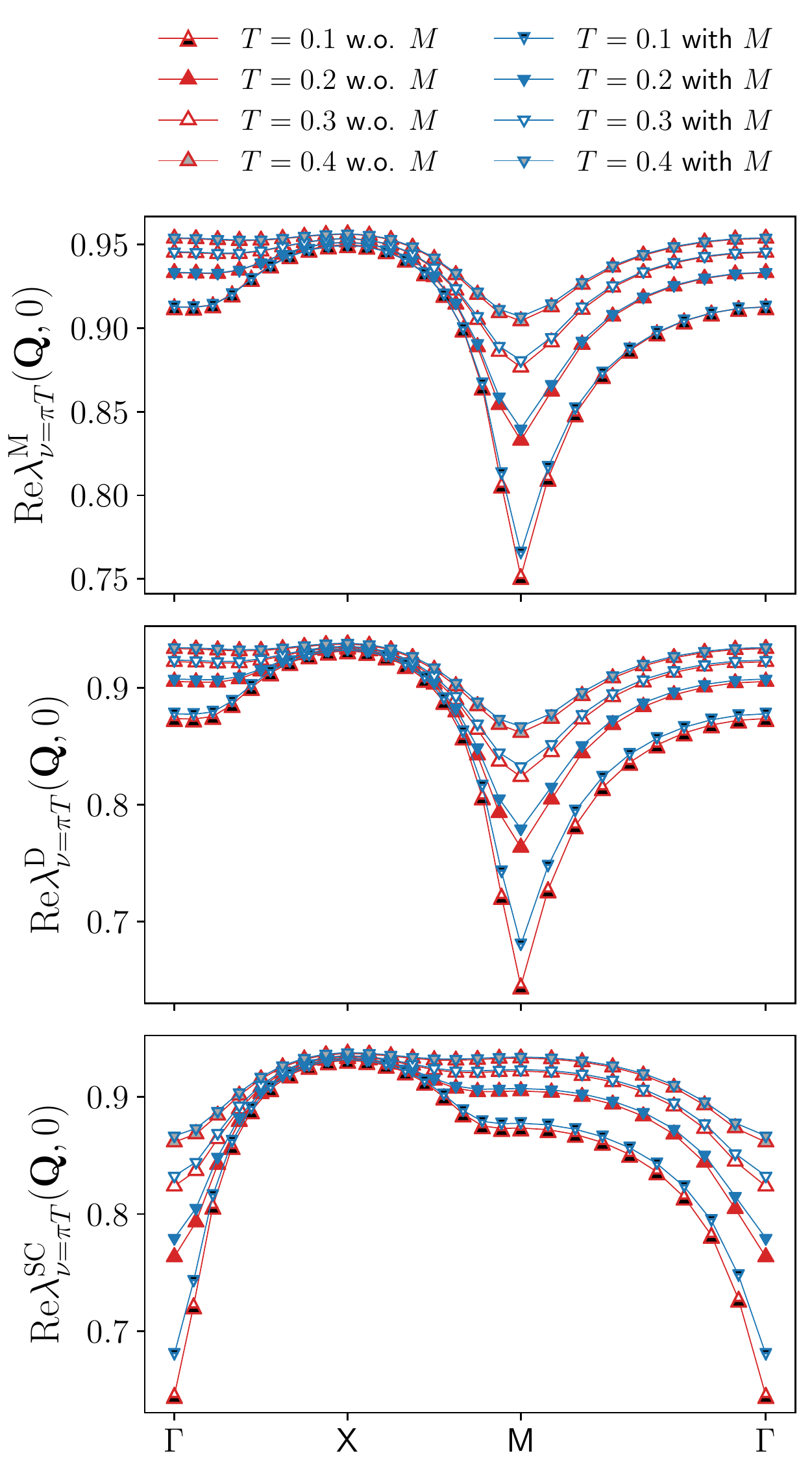}
    \caption{Momentum dependence of the static Yukawa couplings for the magnetic $\lambda_{\nu=\pi T}^\mathrm{M}(\bQ,0)$, density $\lambda_{\nu=\pi T}^\mathrm{D}(\bQ,0)$, and $s$-wave superconducting $\lambda_{\nu=\pi T}^{\mathrm{SC}}(\bQ,0)$ channels as obtained from the SBE formulation of the fRG with (blue symbols) and without rest function (red symbols), for $U=2$ and different values of the temperature ($t'=0$, $\mu=0$). The deviations increase at lower temperatures, with the relative difference between the results with and without rest function below $2.1\%$ for $T\ge 0.2$ and reaching about $5 \%$ for $T=0.1$ in the density and superconducting channels, for the magnetic one see Fig.~\ref{fig: yukawas M vs Q half filling Vary T}.}
    \label{fig: yukawas half filling}
\end{figure}

\begin{figure}[t]
    \centering
    \includegraphics[width = 0.48\textwidth]{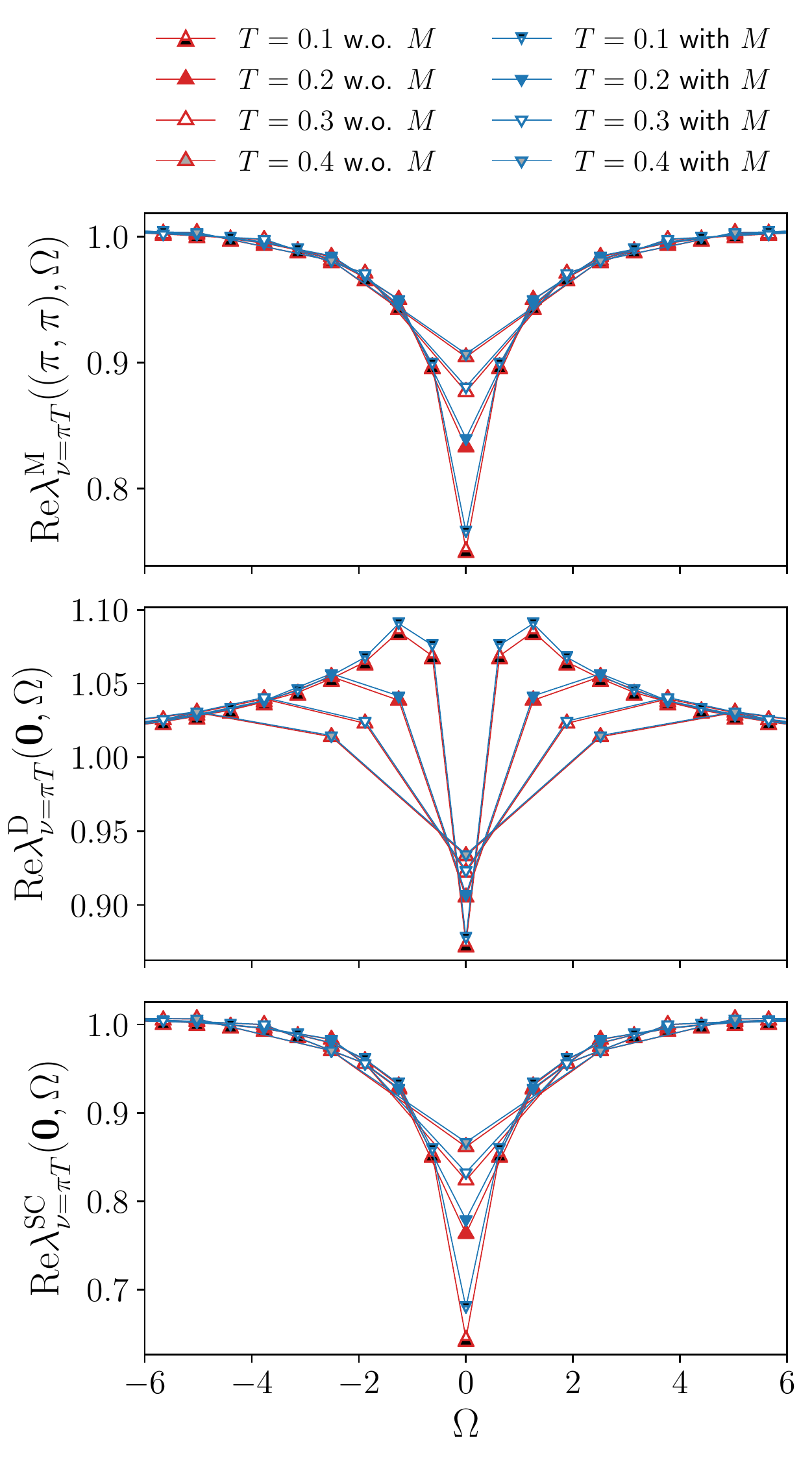}
    \caption{Frequency dependence of the Yukawa couplings for the magnetic $\lambda_{\nu=\pi T}^\mathrm{M}((\pi,\pi),\Omega)$, density $\lambda_{\nu=\pi T}^\mathrm{D}((0,0),\Omega)$, and $s$-wave superconducting $\lambda_{\nu=\pi T}^{\mathrm{SC}}((0,0),\Omega)$ channels for the same parameters as in Fig.~\ref{fig: yukawas half filling}.
    }
    \label{fig: yukawas freq half filling}
\end{figure}

\begin{figure*}[t]
    \centering
    \includegraphics[width = 1.\textwidth]{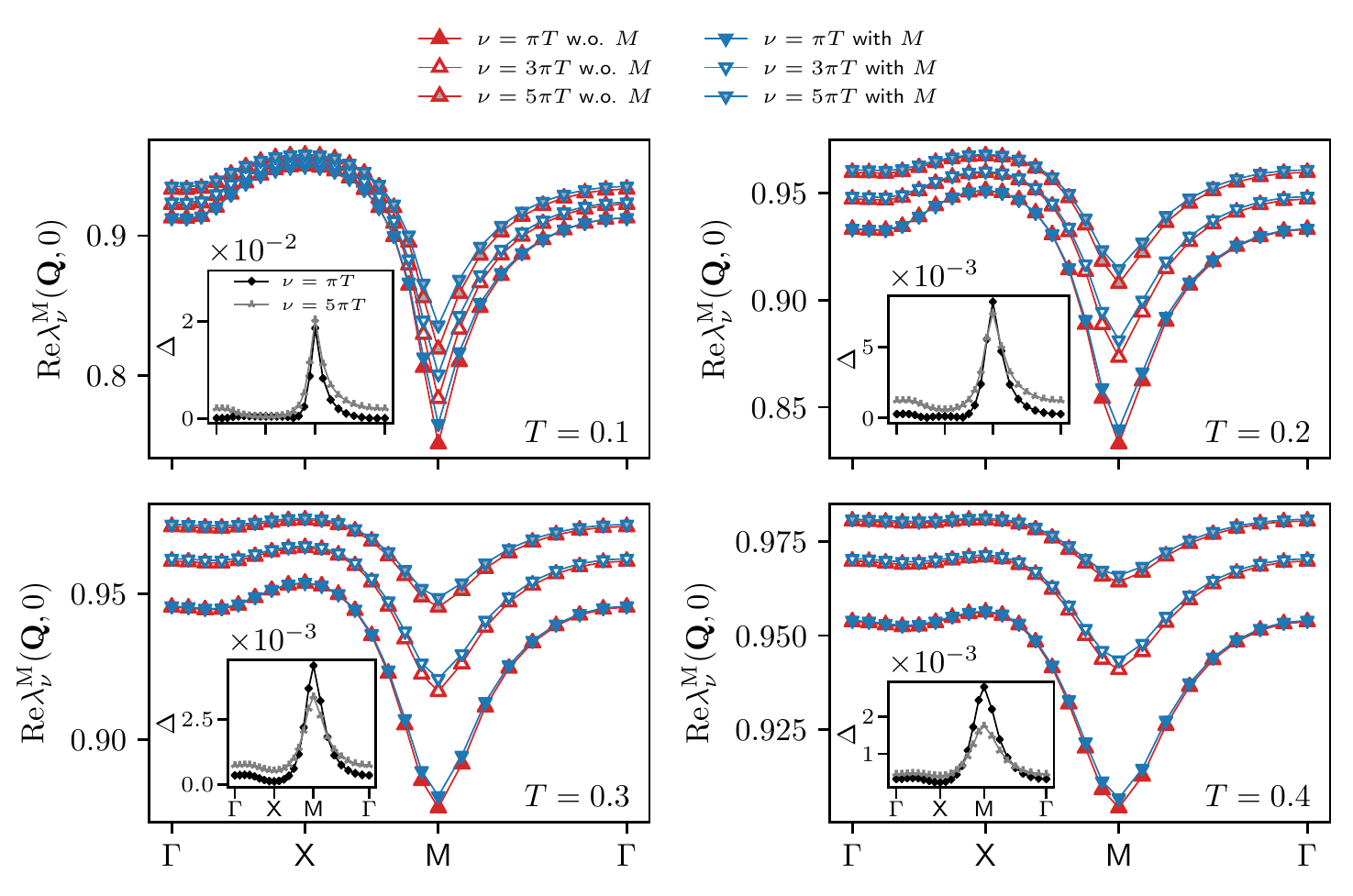}
    \caption{Magnetic channel of the static Yukawa coupling $\lambda_{\nu}^\mathrm{M}(\bQ,0)$ for the same parameters as in Fig.~\ref{fig: yukawas half filling} and different fermionic frequencies. The relative difference $\Delta$ between the results with and without rest function is shown in the insets.}
    \label{fig: yukawas M vs Q half filling Vary T}
\end{figure*}

The results for the Yukawa couplings $\lambda^\mathrm{X}$ (see also Refs.~\cite{VanLoon2018,Krien2019_II,Harkov2021,Harkov2021a}) are shown in Figs.~\ref{fig: yukawas half filling}, \ref{fig: yukawas freq half filling} and \ref{fig: yukawas M vs Q half filling Vary T} for $U=2$ and different values of the temperature. More specifically, Figs.~\ref{fig: yukawas half filling} and~\ref{fig: yukawas freq half filling} display the (bosonic) momentum and frequency dependence of the real parts in the magnetic, density, and superconducting channels respectively. Note that at half filling $\lambda^{\mathrm{SC}}_\nu(\bQ,\Omega)=\lambda^\mathrm{D}_\nu(\bQ+(\pi,\pi),\Omega)$ for the $s$-wave components. The overall reduction with respect to the bare initial value of the Yukawa couplings due to Kanamori screening \cite{Krien2020} shows that their flow can not be neglected (see Appendix~\ref{sec: app} for the effects of the approximation $\lambda^\mathrm{X}=1$). The observed structures become increasingly sharp at low temperature, where also the difference between the results obtained with and without rest function becomes larger. On a quantitative level, the relative error amounts to $5.2\%$ at $T=0.1$, but does not exceed $2.1\%$ for $T\ge 0.2$ in the reported momentum and frequency ranges. The overall agreement in both Figs.~\ref{fig: yukawas half filling} and~\ref{fig: yukawas freq half filling} is very good, demonstrating that the contribution of the rest functions to the Yukawa couplings is minimal in all channels for the smallest fermionic frequency $\nu=\pi T$. We now address the question whether this observation holds also for larger frequencies, see Fig.~\ref{fig: yukawas M vs Q half filling Vary T} where we focus on the dominant magnetic channel. The results for the static (i.e. evaluated at $\Omega=0$) Yukawa couplings at the different temperatures exhibit a moderate dependence on the fermionic frequency $\nu$, featuring an overall slight suppression of the asymptotic value $\lambda^\mathrm{X}\sim 1$. Such a reduction is due to the metallic screening effects \cite{Toschi2007,Katanin2009} in the proximity of the Fermi level. This behavior is observed also in the other channels (not shown). Moreover, Fig.~\ref{fig: yukawas M vs Q half filling Vary T} also shows that the contribution of the rest function remains negligible even for $\nu >\pi T$.

We therefore conclude that neglecting the rest functions leads to quantitatively marginal differences in the Yukawa couplings, for the considered parameter regime at half filling. We will now investigate if it is the case also for the computation of physical observables.

\begin{figure}[t]
    \centering
    \includegraphics[width = 0.48\textwidth]{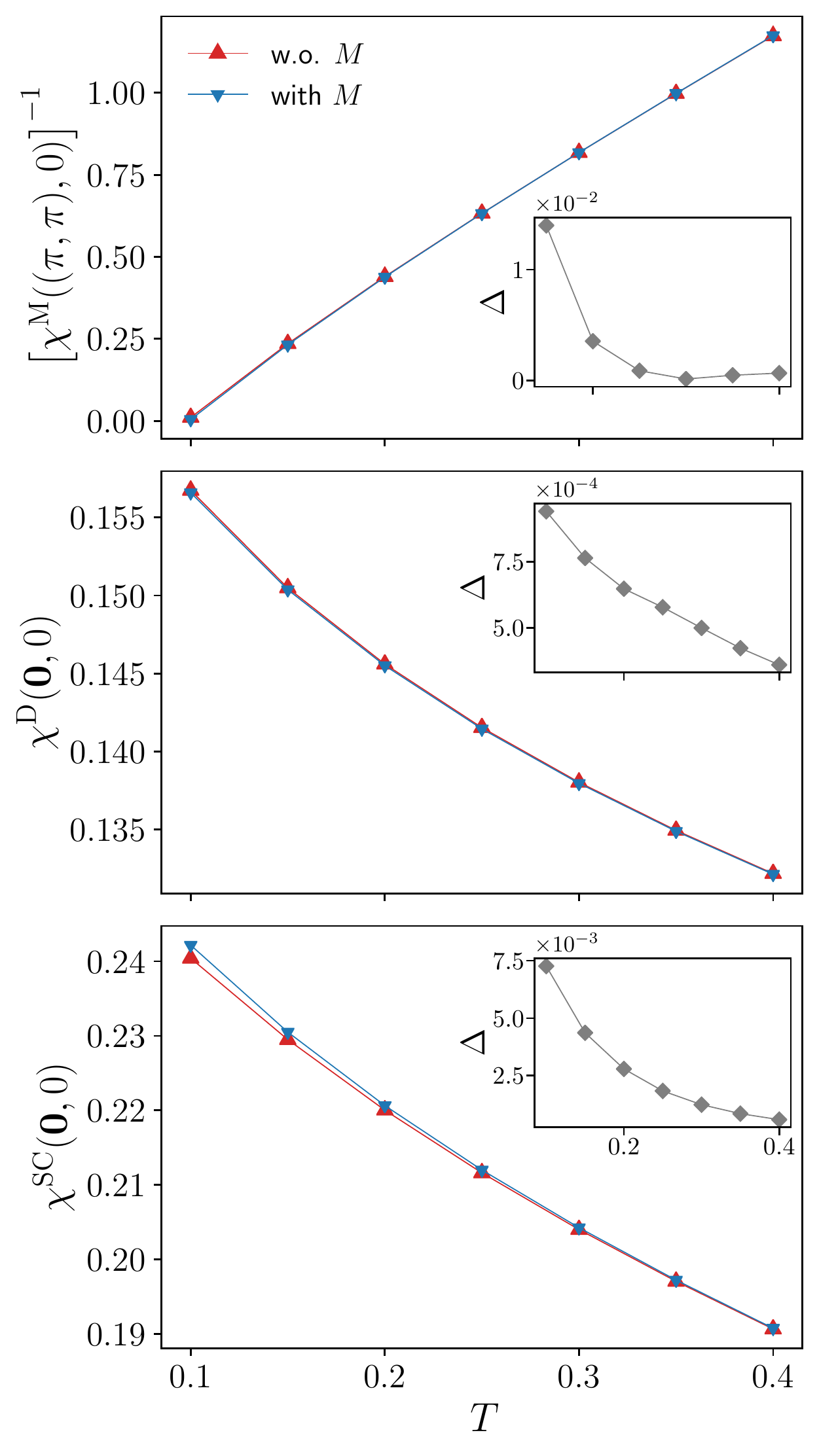}
    \caption{Magnetic $\chi^\mathrm{M}((\pi,\pi),0)$, density $\chi^\mathrm{D}((0,0),0)$, and $s$-wave superconducting $\chi^\mathrm{SC}((0,0),0)$ static susceptibilities as obtained from the SBE formulation of the fRG with and without rest function as a function of $T$, for $U=2$ ($t'=0$, $\mu=0$). The relative difference $\Delta$ between results with and without rest function  displayed in the insets is below $1\%$ except for $T<0.2$ in the magnetic channel. The corresponding results as a function of $U$ for $T=0.15$ are reported in Fig.~\ref{fig: chi U} in the Appendix~\ref{sec: app}.}
    \label{fig: chi T}
\end{figure}

\begin{figure}[t]
    \centering
    \includegraphics[width = 0.48\textwidth]{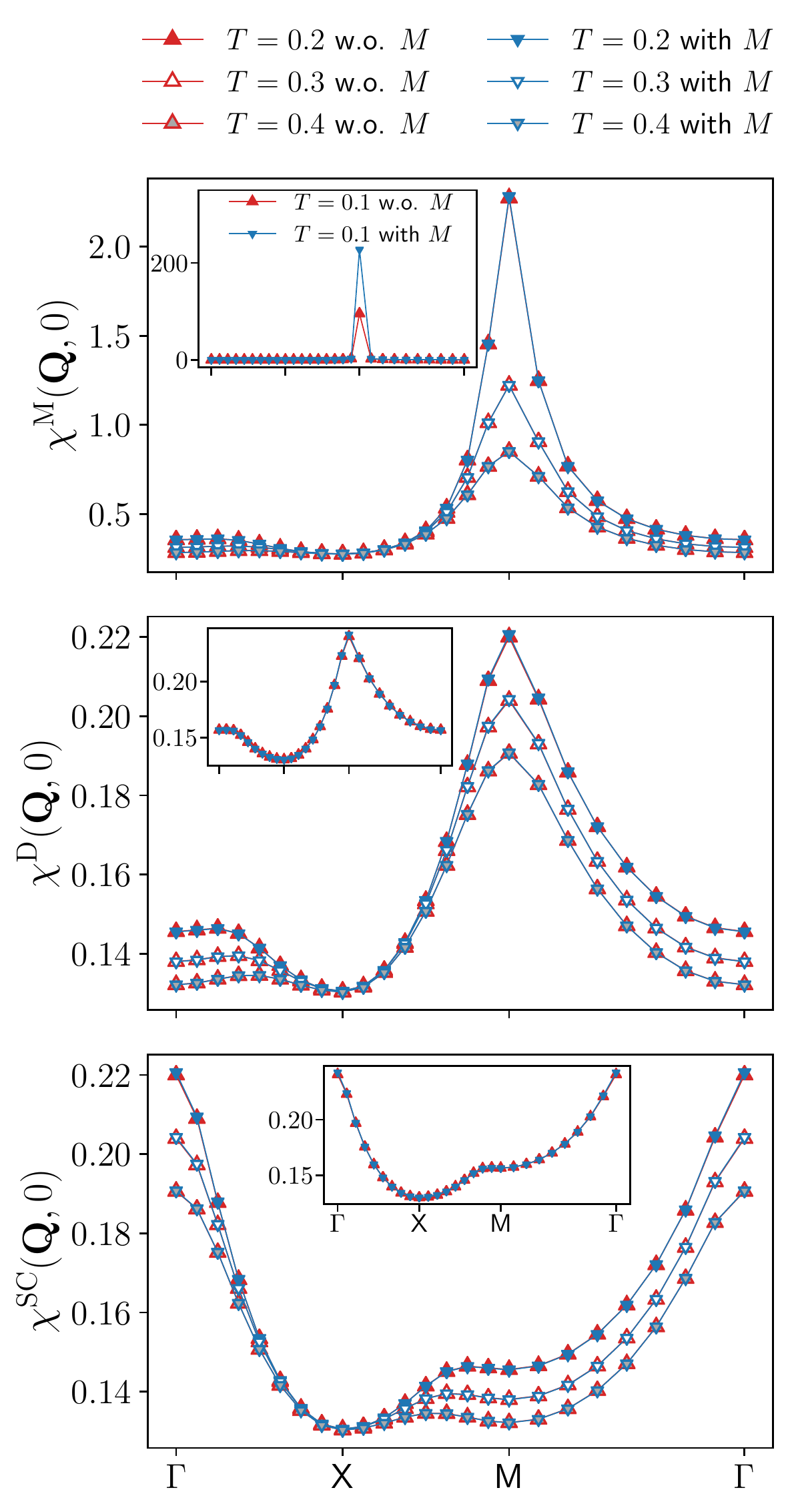}
    \caption{Magnetic $\chi^\mathrm{M}(\bQ,0)$, density $\chi^\mathrm{D}(\bQ,0)$, and $s$-wave superconducting $\chi^{\mathrm{SC}}(\bQ,0)$ static susceptibilities for $U=2$ and different values of the temperature ($t'=0$, $\mu=0$). The results for $T=0.1$ are shown in the insets. The relative differences between the results with and without rest function is below $1\%$  (see also the insets of Fig.~\ref{fig: chi T}), with the exception of the magnetic channel for $T=0.1$.}
    \label{fig: chi momentum}
\end{figure}

\subsection{Susceptibilities}

We here study the impact of the rest functions $M^\mathrm{X}$ on our numerical results for the physical observables, more specifically for the susceptibilities $\chi^\mathrm{X}$ (results for the self-energy are provided in the Appendix~\ref{sec: app}). We start by presenting their evolution with the temperature, displayed in Fig.~\ref{fig: chi T} for $U=2$. As for the previous figures, the static susceptibilities, evaluated at the relevant momentum values $\bf Q$ in the magnetic, density, and $s$-wave superconducting channels, are compared to the ones obtained without the SBE rest functions $M^\mathrm{X}$. From the relative difference shown in the insets, we see that the contribution of the rest function increases with the inverse temperature. At the same time, the absolute values are very small leading to a marginal effect of the rest functions except for the magnetic susceptibility at $T=0.1$. Before discussing this point, we remark that in the density and superconducting channels the importance of the rest function is significantly smaller than in the dominant magnetic channel. A similar behavior is observed also for the interaction dependence of the susceptibilities (see Fig.~\ref{fig: chi U} in the Appendix~\ref{sec: app}). As expected, the contribution of the rest function increases with $U$, but remains marginal in all channels for $T>0.1$. We also note that the susceptibility of the dominant magnetic channel $\chi^\mathrm{M}$, which is mostly driven by vertex corrections, significantly increases with the interaction, whereas the bubble dominated subleading susceptibilities, whose largest contribution originates from the bubble term, decrease with $U$ since the growing AF fluctuations lead to stronger damping via the electronic self-energy. The increase at lower temperatures is due to the growing AF fluctuations for the magnetic channel and a Fermi-liquid like behavior for the subleading channels.

We now turn to the large contribution of the rest function obtained for the magnetic channel at low temperatures, with a deviation of $\Delta > 50 \%$ between the results with and without rest function. In the inverse susceptibility displayed in the upper main panel of Fig.~\ref{fig: chi T} this is not so clearly visible because of the small absolute values. At the same time, the almost vanishing inverse susceptibility at $T=0.1$ indicates the proximity to a divergence in the magnetic channel\footnote{To access temperatures below the pseudo-critical temperature, one needs to allow for the formation of an order parameter in the system and therefore to include anomalous components of the vertices \cite{Metzner12}.}. The diverging one-loop flow in correspondence of a finite temperature (or interaction) marks the pseudo-critical transition towards a magnetic instability. Including higher loop orders within the multiloop fRG extension~\cite{Kugler2018_I,Tagliavini2019,Hille2020,Chalupa2022} would reduce the strong AF fluctuations due to the stronger channel interference. In particular, the resummation to infinite loop order in the parquet approximation would recover the Mermin-Wagner theorem. Here, we observe that sizable differences for the computation with and without the rest functions $M^\mathrm{X}$ arise only in the dominant magnetic susceptibility close to the pseudo-critical AF transition. Away from it, the tiny differences justify the approximation to neglect the rest function.

The appearance of the pseudo-critical transition characteristic of the one-loop fRG approximation is further illustrated in Fig.~\ref{fig: chi momentum}, more specifically in the inset of the panel reporting the magnetic susceptibility. In this figure, we show the momentum dependence of the static magnetic, density, and $s$-wave superconducting susceptibilities as obtained from the inversion of Eq.~\eqref{eq: D} for the screened interactions, for $U=2$ and different temperatures (the corresponding results for the frequency dependence are provided in the Appendix~\ref{sec: app}). The magnetic susceptibility exhibits a pronounced peak around momentum $\bs{Q}=(\pi,\pi)$ indicating strong AF fluctuations. Except for $\chi^\mathrm{M}({\bf Q},0)$ at $T=0.1$, the results of the computation with and without the rest function for the dominating magnetic channel as well as for the subleading channels related by $\chi^\mathrm{SC}(\bQ,\Omega)=\chi^\mathrm{D}(\bQ+(\pi,\pi),\Omega)$ at particle-hole symmetry present an excellent quantitative agreement, with the largest deviation in correspondence of the AF wave vector. The relative difference for $\chi^\mathrm{M}({\bf Q},0)$ is below $1\%$ (except for $T=0.1$) and as a consequence, the inclusion of the rest function only sightly affects the pseudo-critical temperature~\cite{Bonetti2022}. For $\chi^\mathrm{D}({\bf Q},0)$ and $\chi^\mathrm{SC}({\bf Q},0)$, the relative difference is even smaller, below $0.1\%$ for all values of the temperature.

\begin{figure*}[t]
    \centering
    \includegraphics[width = .7 \textwidth]{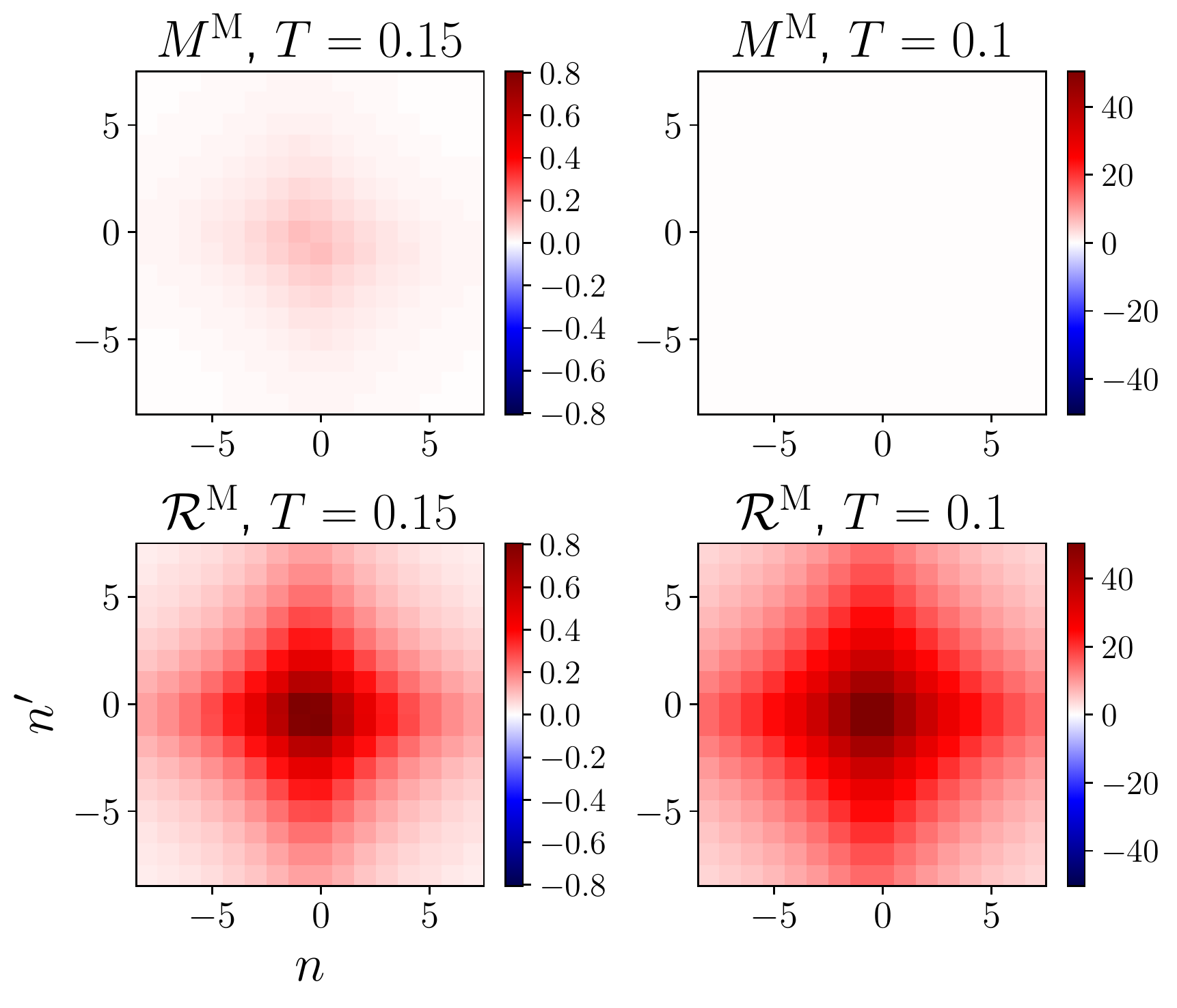}
    \caption{Rest functions $M^\mathrm{M}_{\nu\nu'}((\pi,\pi),0)$ (upper panels) and $\mathcal{R}^\mathrm{M}_{\nu\nu'}((\pi,\pi),0)$ (lower panels) in the magnetic channel, as obtained respectively from the SBE and the conventional fermionic fRG formulations, with $n$ and $n^{\prime}$ labeling the fermionic Matsubara frequencies according to $\nu^{(\prime)}=(2n^{(\prime)}+1)\pi T$, for $U=2$ and $T=0.1$ and $T=0.15$ ($t'=0$, $\mu=0$). Note that the SBE rest function almost vanishes in comparison (for the absolute values see Figs.~\ref{fig: rest functions} and \ref{fig: rest functions 2} provided in the Appendix~\ref{sec: app}, where also the contributions of the other channels are displayed).}
    \label{fig: rest functions Main}
\end{figure*}

$T=0.1$ is very close to the pseudo-critical transition temperature. We note that the associated divergence in the susceptibility is reflected in the corresponding screened interaction but \emph{not} in the Yukawa coupling, see Figs.~\ref{fig: yukawas half filling}-\ref{fig: yukawas M vs Q half filling Vary T}. The multiboson exchange processes associated to spin fluctuations at low temperatures, appear to induce only a marginal increase of the Yukawa couplings. While the rest function $M^\mathrm{M}$ too does not diverge (see Fig.~\ref{fig: rest functions Main}), its influence on $\chi^\mathrm{M}$ may become nonnegligible. This has to be contrasted to the conventional fermionic fRG formulation, where all objects diverge in proximity of the pseudo-critical transition. Since resolving divergent structures is computationally expensive, the latter requires a higher numerical effort with respect to the SBE implementation.

\begin{figure}[t]
    \centering
    \includegraphics[width = 0.48\textwidth]{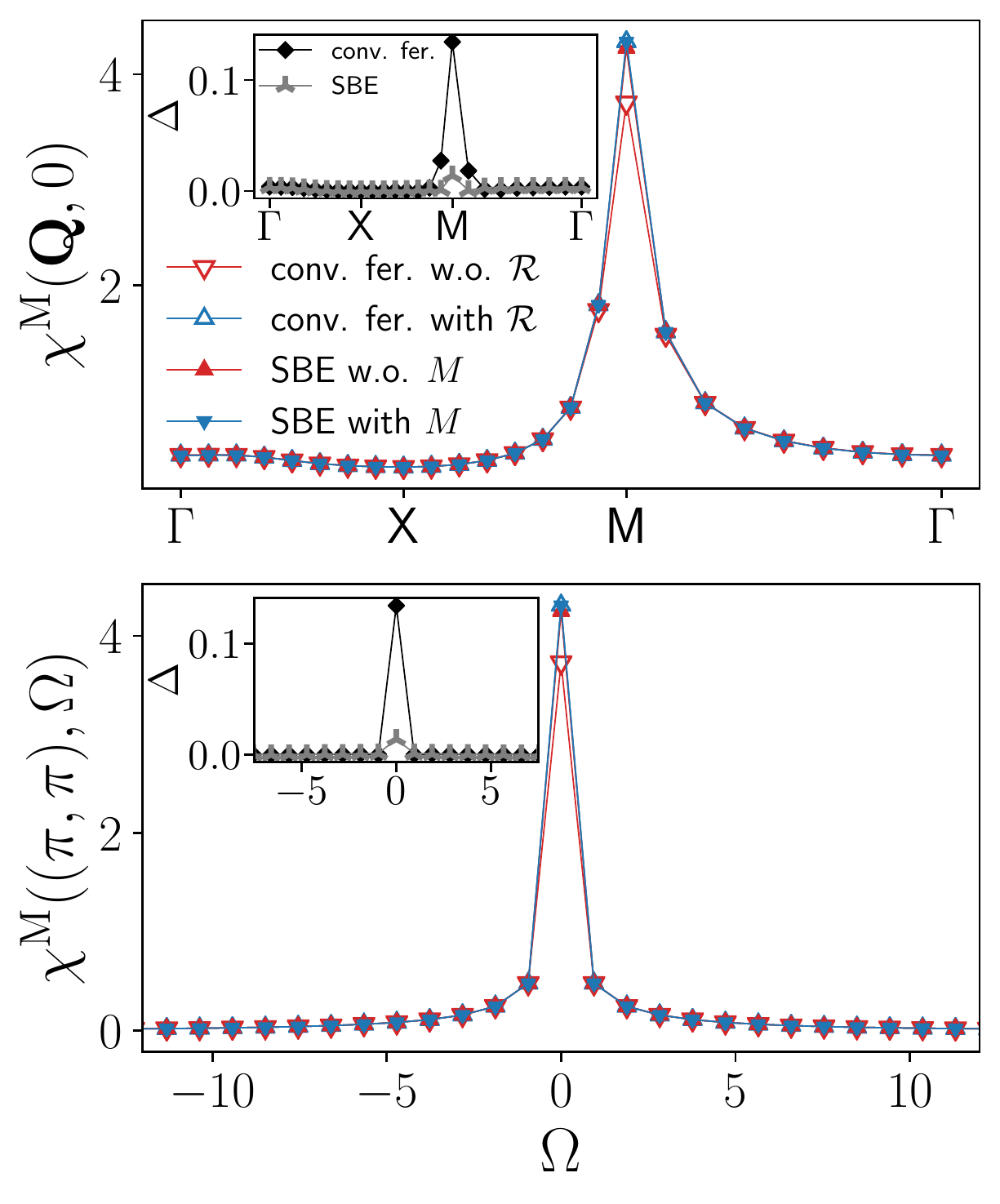}
    \caption{Bosonic momentum and frequency dependence of the magnetic susceptibility $\chi^\mathrm{M}$ as obtained from the SBE and conventional fermionic fRG formulations with and without rest function, for $U=2$ and $T=0.15$ ($t'=0$, $\mu=0$). The insets show the relative difference $\Delta$ between results with and without rest function.}
    \label{fig: chi momentum 2}
\end{figure}

The convenience of the SBE formulation emerges also from the direct comparison with data obtained from the conventional fermionic fRG. Figure~\ref{fig: chi momentum 2} provides the results for the leading magnetic susceptibility as a function of both the (bosonic) momentum and frequency, for $U=2$ and $T=0.15$. At the maximum at $\bQ=(\pi,\pi)$ and $\Omega=0$, the relative difference between the results with and without rest function reported in the insets is about $10$ times smaller in the SBE decomposition with respect to the one in the conventional fermionic formulation, while away from it the contribution of both (SBE and conventional fermionic fRG) rest functions rapidly decays. In particular, the effect of the SBE rest function is marginal in the whole momentum and frequency range, in contrast to the conventional fRG implementation. In the following, we will provide an explanation for this discrepancy by analysing the respective rest functions in more detail.

\subsection{Rest functions}

Fig.~\ref{fig: rest functions Main} shows the fermionic frequency dependence of the rest functions $M^\mathrm{M}$ and $\mathcal{R}^\mathrm{M}$ in the magnetic channel, evaluated at $\bQ=(\pi,\pi)$ and $\Omega=0$ for $T=0.15$ as well as $T=0.1$ in proximity of the AF pseudo-critical transition (plots including also the density and superconducting channels are provided in Figs.~\ref{fig: rest functions} and~\ref{fig: rest functions 2} in the Appendix~\ref{sec: app}). It highlights a key feature of the rest function:  its characteristic decay to zero at large frequencies (which is particularly pronounced for larger values of the interaction~\cite{Bonetti2022}). This implies that the high-frequency asymptotics of the two-particle vertex functions is fully captured by the screened interactions $w^\mathrm{X}$ and the Yukawa couplings $\lambda^\mathrm{X}$. 

We also illustrate from Fig.~\ref{fig: rest functions Main} a basic symmetry property of both SBE and conventional fermionic rest functions, which is not inherent to the magnetic channel. Namely, the rest functions can be decomposed as
\begin{equation}
 M^{\mathrm{X}}_{k k'}(Q) = M^{+,\mathrm{X}}_{k k'}(Q)+M^{-,\mathrm{X}}_{k k'}(Q) ,
\end{equation}
with $M^{+/-,\mathrm{X}}$ being symmetric/antisymmetric with respect to the inversion of a single fermionic frequency, i.e.
\begin{equation}
M^{\pm,\mathrm{X}}_{(\bs{k},\nu)(\bs{k}',\nu')}(Q) = \frac{1}{2}\big[M^{\mathrm{X}}_{(\bs{k},\nu)(\bs{k}',\nu')}(Q) 
\pm M^{\mathrm{X}}_{(\bs{k},-\nu)(\bs{k}',\nu')}(Q)\big].
\end{equation}
In most situations, $M^{+,\mathrm{X}}$ is the dominant contribution, which is clearly the case for $M^{\mathrm{M}}$ in Fig.~\ref{fig: rest functions Main}.

Comparing the results for the SBE and the conventional fermionic decompositions, we observe a marked difference in the absolute values of the rest functions. For $T=0.15$ the SBE rest function is almost an order of magnitude smaller, with $M_{\nu\nu'}((\pi,\pi),0)\ll\mathcal{R}_{\nu\nu'}((\pi,\pi),0)$ for the maximal values. As seen in Fig.~\ref{fig: chi momentum 2}, it is in fact essential to include the contribution of the rest function  in the conventional fermionic fRG, while in the SBE formulation it leads only to minor corrections. This discrepancy is further enhanced as we lower the temperature down to $T=0.1$, due to the dramatic increase of   $\mathcal{R}^\mathrm{M}$ in proximity of the pseudo-critical transition  (in contrast, $M^\mathrm{M}$ is of the same order as for $T=0.15$, see Appendix~\ref{sec: app}). The difference between $M^\mathrm{M}$ and $\mathcal{R}^\mathrm{M}$ can be traced back to the (magnetic) screened interaction: according to Eq.~\eqref{eq: R_SBE vs R_asym}, the increase of $\mathcal{R}_{\nu\nu'}^{\mathrm{M}}((\pi,\pi),0)$ is absorbed by its $U$-reducible part, specifically by the growing screened interaction, while the SBE rest function stays small even in the vicinity of the pseudo-critical transition. Also this regime can thus be reliably described by the SBE scheme without rest function, which significantly reduces the numerical effort of the fRG flow. For the considered parameter regime, the gain in the run times for the computation without SBE rest function is parameter dependent ($\sim 3-4$ for $T=0.1-0.15$), but most importantly the lower number of required momentum and frequency variables entails a better scaling.

\begin{figure*}[t]
    \centering
    \includegraphics[width = 1.\textwidth]{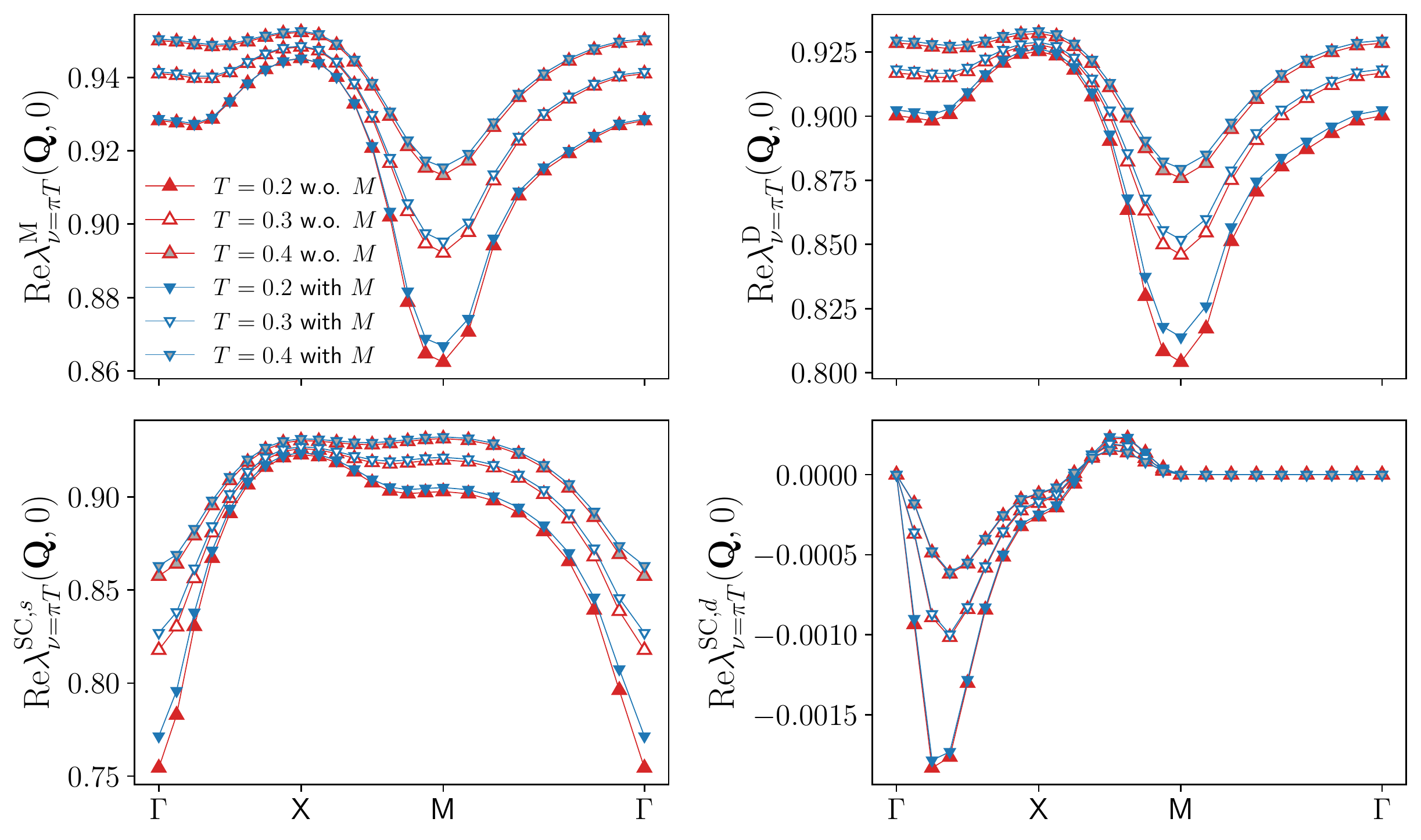}
    \caption{Momentum dependence of the static Yukawa couplings for the magnetic $\lambda_{\nu=\pi T}^\mathrm{M}(\bQ,0)$, density $\lambda_{\nu=\pi T}^\mathrm{D}(\bQ,0)$, and both $\lambda_{\nu=\pi T}^{\mathrm{SC},s}(\bQ,0)$ and $\lambda_{\nu=\pi T}^{\mathrm{SC},d}(\bQ,0)$ superconducting channels as obtained from the SBE formulation of the fRG with (blue symbols) and without rest function (red symbols), for $U=2$, $t^\prime=-0.2$, $\mu=4t'$, and different values of the temperature. At the end of the flow, the filling equals $0.44$ for all temperatures (with $0.5$ corresponding to half filling), with or without rest function.}
    \label{fig: yukawas doping}
\end{figure*}

Summarizing, we have shown that in the SBE formulation of the fRG the contribution of the rest function is marginal as long as the one-loop flow does not diverge. In proximity of the divergence, it becomes relevant on a quantitative level, but does not qualitatively affect the physics. Differently to the conventional fermionic fRG implementation, the screened interactions appear to effectively absorb the divergence. Thus, the $U$-reducible part of the two-particle vertex captures the physically relevant information near the pseudo-critical transition. The physical susceptibilities (as well as the self-energy, see Appendix~\ref{sec: app}) can hence be reliably determined by the flow of the screened interactions and Yukawa couplings, without including the rest function. This puts forward the SBE-based fRG as a computationally more efficient alternative to the conventional fermionic description. In the following section we will examine whether this is also the case at finite doping, where superconducting correlations are expected to play a more prominent role.

\section{Results at finite doping}
\label{sec:results doped regime}

In order to illustrate the validity of the SBE formulation in a physically more relevant parameter regime than the perfect particle-hole symmetric case, we present also fRG results at finite doping. Specifically, we consider a next-nearest neighbor hopping close to van Hove filling at $\mu=4t'$ (due to the flow of the self-energy, the initial value of the chemical potential is renormalized during the flow), for which we expect the magnetic fluctuations to be suppressed in favor of superconducting correlations. These are taken into account by the flow of the $d$-wave components (and their interplay with the $s$-wave components) in the different channels. As in Section~\ref{sec:results half filling}, we will analyse the effects of the rest functions on the Yukawa couplings and physical susceptibilities, including also a comparison to the conventional fermionic formulation.

\subsection{Yukawa couplings}

\begin{figure}[t]
    \centering
    \includegraphics[width = 0.48\textwidth]{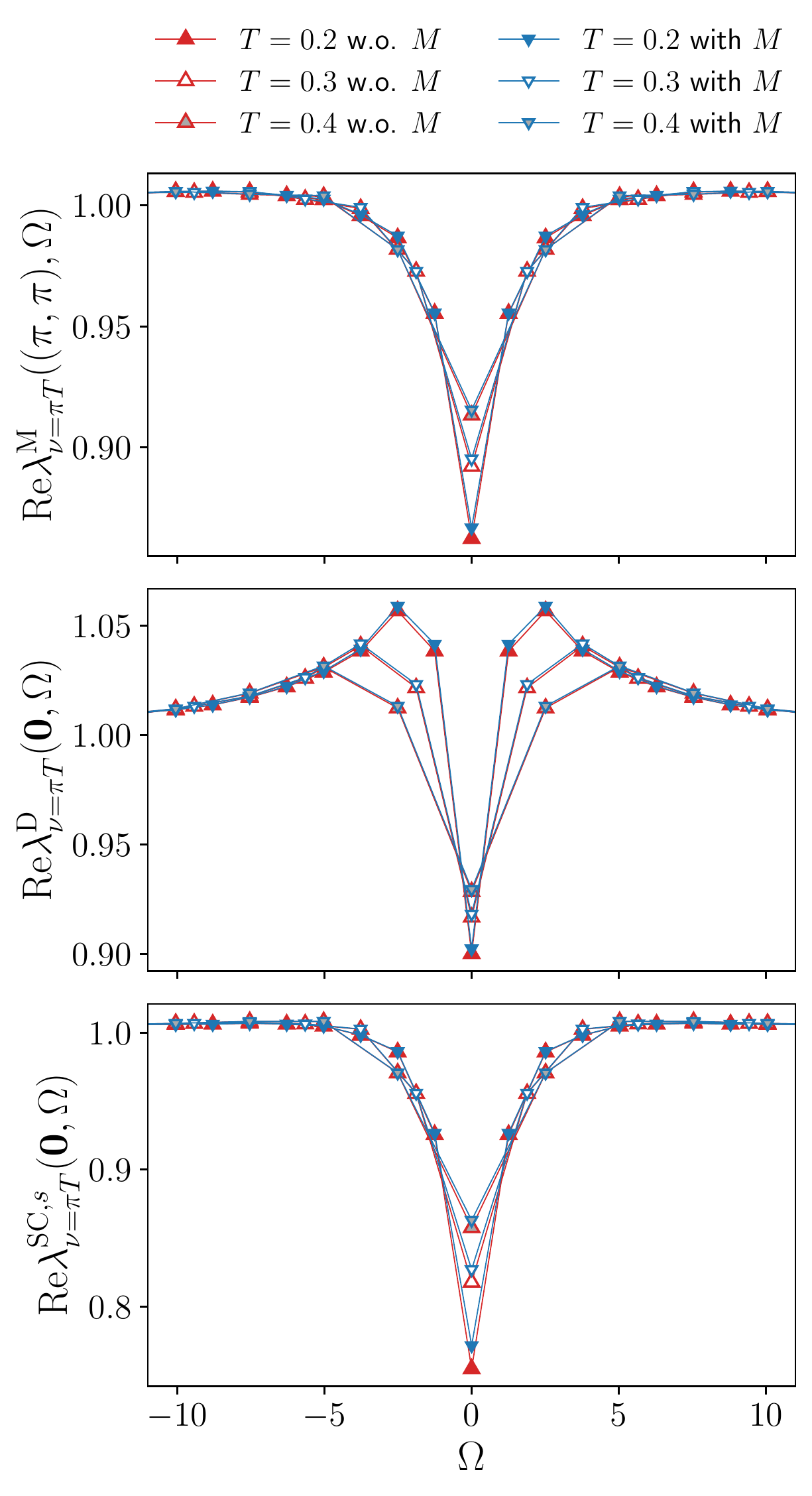}
    \caption{Frequency dependence of the Yukawa couplings for the magnetic $\lambda_{\nu=\pi T}^\mathrm{M}((\pi,\pi),\Omega)$, density $\lambda_{\nu=\pi T}^\mathrm{D}((0,0),\Omega)$, and $s$-wave superconducting $\lambda_{\nu=\pi T}^{\mathrm{SC},s}((0,0),\Omega)$ channels for the same parameters as in Fig.~\ref{fig: yukawas doping}. Note that $\lambda^{\mathrm{SC},d}(\bs{Q}=\bs{0},\Omega)=0$.
    }
    \label{fig: yukawas freq doping}
\end{figure}

The results for the Yukawa couplings in all relevant channels are shown in Figs.~\ref{fig: yukawas doping} and~\ref{fig: yukawas freq doping}, illustrating the (bosonic) momentum and frequency dependence of the real parts evaluated at the lowest fermionic Matsubara frequency $\nu=\pi T$, for $U=2$, $t'=-0.2$, and different temperatures. The general shapes in the magnetic, density and $s$-wave superconducting channels resemble those of Figs.~\ref{fig: yukawas half filling} and~\ref{fig: yukawas freq half filling} at half filling. For the momentum dependence, $\lambda^{\mathrm{M}}$ and $\lambda^{\mathrm{D}}$ are suppressed in correspondence of the magnetic ordering wavevector $\bQ=(\pi,\pi)$, whereas $\lambda^{\mathrm{SC},s}$ falls off near the $\mathrm{\Gamma}$-point. For the frequency dependence, the reduction in the low-frequency regime results from the electronic screening. Lowering the temperature, these features become sharper.

An important difference to the previous study at half filling consists in the finite $d$-wave component $\lambda^{\mathrm{SC},d}$, which exhibits a slight dip between the $\mathrm{\Gamma}$- and X-points (see Fig.~\ref{fig: yukawas doping}). We further note that $\lambda^{\mathrm{SC},d}(\bs{Q}=\bs{0},\Omega)=0$ for all bosonic frequencies due to the vanishing of all mixed form-factor bubbles at the $\mathrm{\Gamma}$-point in a framework involving only $s$- and $d$-wave form factors~\cite{Heinzelmann22}. This can be understood diagrammatically: the $U$-reducible part of the two-particle vertex in the $d$-wave superconducting channel involves the product
\begin{equation}
\begin{gathered}
\begin{fmffile}{Diagrams/DiagLambdaSCd}
\begin{fmfgraph*}(140,80)
\fmfleft{i0,i1,i2,i3,i4,i5,i6}
\fmfright{o0,o1,o2,o3,o4,o5,o6}
\fmf{fermion,tension=2.5}{i5,vUpL}
\fmf{phantom,tension=0.5}{o5,vUpL}
\fmf{fermion,tension=2.5}{i1,vDownL}
\fmf{phantom,tension=0.5}{o1,vDownL}
\fmf{phantom,tension=1.5}{i3,vRight}
\fmf{phantom,tension=1.5}{o3,vRight}
\fmf{phantom,tension=0.2}{i3,vRight2}
\fmf{phantom,tension=1.5}{o3,vRight2}
\fmf{plain,tension=0.}{vDownL,vUpL}
\fmf{wiggly,tension=0.2}{vRight,vRight2}
\fmf{plain,tension=0.}{vDownL,vRight}
\fmf{plain,tension=0.}{vUpL,vRight}
\fmfv{label.angle=180,label.dist=0.5cm,label=$\lambda^{\mathrm{SC},,d}$}{vRight}
\end{fmfgraph*}
\end{fmffile}
\end{gathered}
\label{diag1}
\end{equation}
between the Yukawa coupling $\lambda^{\mathrm{SC},d}$ and the screened interaction $w^\mathrm{SC}$. Since the latter has only $s$-wave components, all contributions to~\eqref{diag1} contain mixed bubbles with at least one propagator connecting different form factors, such as
\begin{equation}
\begin{gathered}
\begin{fmffile}{Diagrams/DiagLambdaSCdmixedbubble}
\begin{fmfgraph*}(140,80)
\fmfleft{i0,i1,i2,i3,i4,i5,i6}
\fmfright{o0,o1,o2,o3,o4,o5,o6}
\fmf{fermion,tension=2.5}{i5,vUpL}
\fmf{phantom,tension=0.5}{o5,vUpL}
\fmf{fermion,tension=2.5}{i1,vDownL}
\fmf{phantom,tension=0.5}{o1,vDownL}
\fmf{phantom,tension=1.5}{i3,vRight}
\fmf{phantom,tension=1.5}{o3,vRight}
\fmf{phantom,tension=0.2}{i3,vRight2}
\fmf{phantom,tension=1.5}{o3,vRight2}
\fmf{fermion,tension=0.,left=0.3}{vDownL,vUpL}
\fmf{fermion,tension=0.,left=0.3}{vUpL,vDownL}
\fmf{wiggly,tension=0.2}{vRight,vRight2}
\fmf{fermion,tension=0.,right=0.2}{vDownL,vRight}
\fmf{fermion,tension=0.,left=0.2}{vUpL,vRight}
\fmfv{label.angle=75,label.dist=0.15cm,label=$s$}{vRight}
\fmfv{label.angle=90,label.dist=0.11cm,label=$d$}{vUpL}
\fmfv{label.angle=-90,label.dist=0.08cm,label=$d$}{vDownL}
\end{fmfgraph*}
\end{fmffile}
\end{gathered}
\label{diag2}
\end{equation}
where the form factor indices are indicated explicitly and the fermionic propagators $G(k)$ are represented by solid lines. The reason why these mixed bubbles vanish exactly at the $\mathrm{\Gamma}$-point ($\bs{Q}=\bs{0}$) can be illustrated through the bare superconducting susceptibility
\begin{equation}
\begin{split}
 \left[\chi^\mathrm{SC}_{0}\right]_{mn} (\bs{Q},i\Omega) = & \sum_{i\nu} \int d\bs{k} f_{n}^{*}(\bs{k}) f_{m}(\bs{k}) G(\bs{k},i\nu) \\
 & \times G(\bs{Q}-\bs{k},i\Omega-i\nu),
\end{split}
\end{equation}
with $m$ and $n$ form factor indices. This simplest mixed bubble of interest for $m=0$ and $n=1\neq m$ reads (for $\Omega=0$)
\begin{equation}
\begin{split}
 \left[\chi^\mathrm{SC}_{0}\right]_{01} (\bs{Q},0) = & \sum_{i\nu} \int d\bs{k} (\cos(k_x)-\cos(k_y)) G(\bs{k},i\nu) \\
 & \times G(\bs{Q}-\bs{k},-i\nu),
\end{split}
\label{eq:MixedBubblevanish}
\end{equation}
where the $s$-wave $f_{0}(\bs{k})=1$ and $d$-wave $f_{1}(\bs{k})=\cos(k_x)-\cos(k_y)$ form factors have been inserted. At $\bs{Q}=\bs{0}$, the momentum integral in Eq.~\eqref{eq:MixedBubblevanish} vanishes exactly. This reasoning can be straightforwardly generalized to all mixed bubbles involving $s$- and $d$-wave form factors, such as that of~\eqref{diag2}.

Concerning the contribution of the rest function to our results for the Yukawa couplings at finite doping, the relative difference between the results with and without rest function in the magnetic, density and $s$-wave superconducting channels is below $2.3 \%$ in both Figs.~\ref{fig: yukawas doping} and~\ref{fig: yukawas freq doping}. Because of the zeros of $\mathrm{Re}\lambda_{\nu=\pi T}^{\mathrm{SC},d}(\bQ,0)$, for the $d$-wave superconducting channel we consider the absolute difference, which does not exceed $5 \cdot 10^{-5}$ for all temperatures. Hence, the overall contribution of the rest function to the Yukawa couplings is still negligible in all channels.

\begin{figure}[t]
    \centering
    \includegraphics[width = 0.48\textwidth]{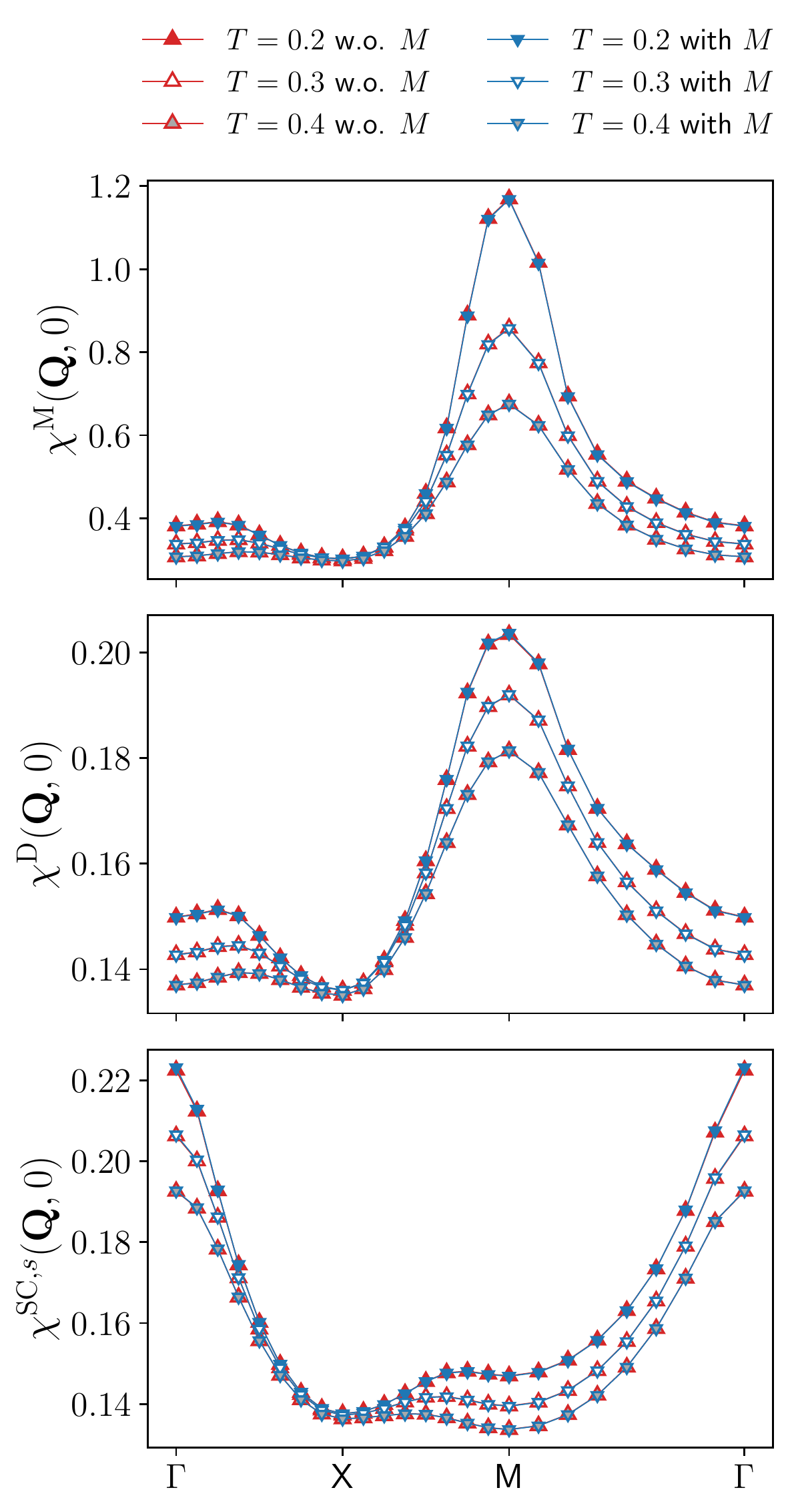}
    \caption{Magnetic $\chi^\mathrm{M}(\bQ,0)$, density $\chi^\mathrm{D}(\bQ,0)$, and $s$-wave superconducting $\chi^{\mathrm{SC},s}(\bQ,0)$ static susceptibilities as obtained from the SBE formulation of the fRG with and without rest function, for the same parameters as in Fig.~\ref{fig: yukawas doping}. The relative difference between the results with and without rest function is below $1\%$ for all temperatures in the magnetic, density and $s$-wave superconducting channels.}
    \label{fig: chi doping U2}
\end{figure}

\begin{figure}[t]
    \centering
    \includegraphics[width = 0.48\textwidth]{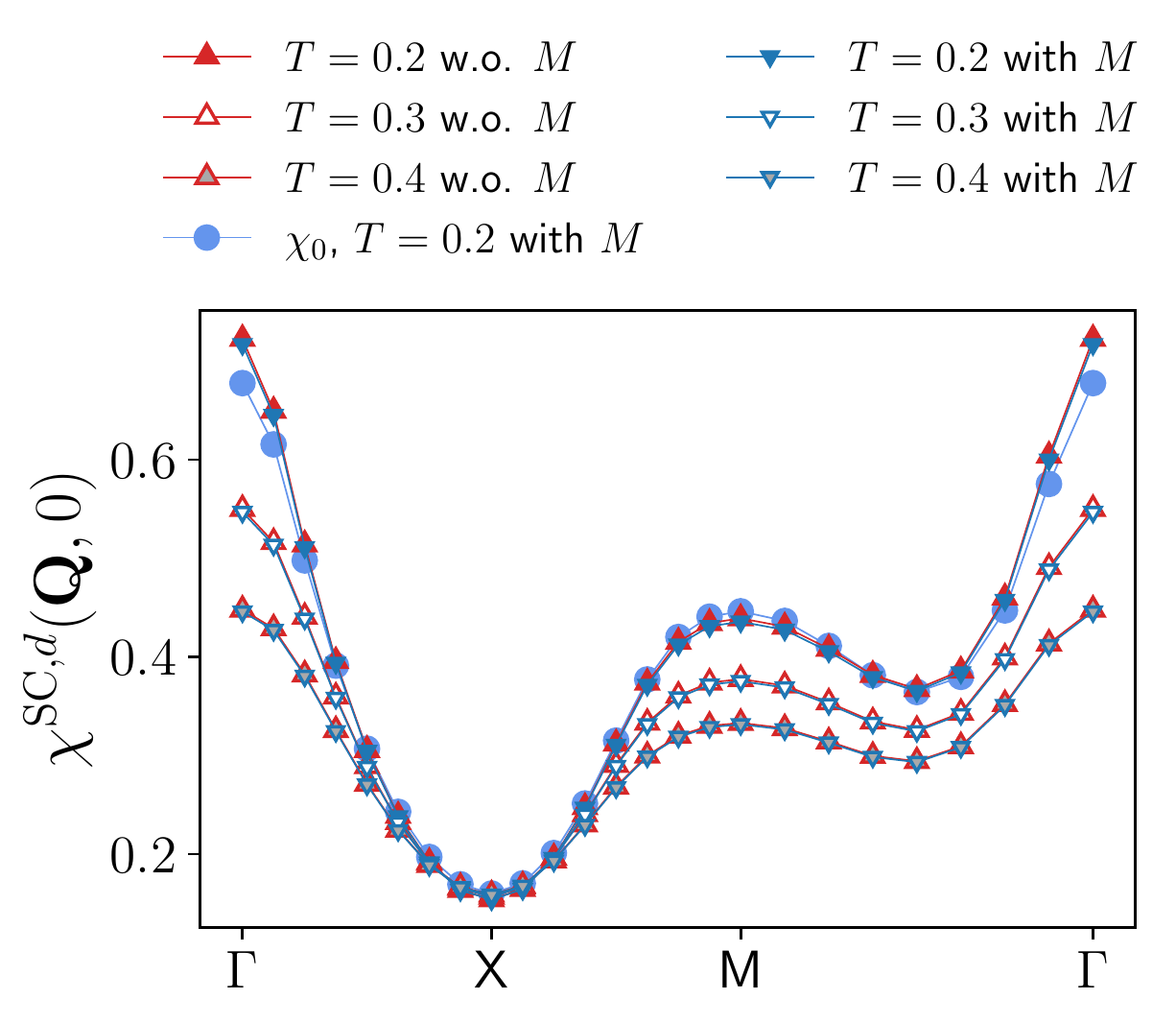}
    \caption{Momentum dependence of the $d$-wave superconducting susceptibility $\chi^{\mathrm{SC},d}$, determined from post-processing with and without rest function, for the same parameters as in Fig.~\ref{fig: yukawas doping}. The relative difference between the results with and without rest function is below $1 \%$ for all temperatures. Additionally, results for the bare bubble contribution $\chi_{0}$ are provided for $T=0.2$. There, the relative difference between the full $d$-wave superconducting susceptibility and its bare bubble contribution reaches $5\%$ at the $\mathrm{\Gamma}$-point (with or without rest function), which decreases for larger temperatures.}
    \label{fig: chi doping U2 bis}
\end{figure}

\subsection{Susceptibilities}

\begin{figure*}[t]
    \centering
    \includegraphics[width = 1.0 \textwidth]{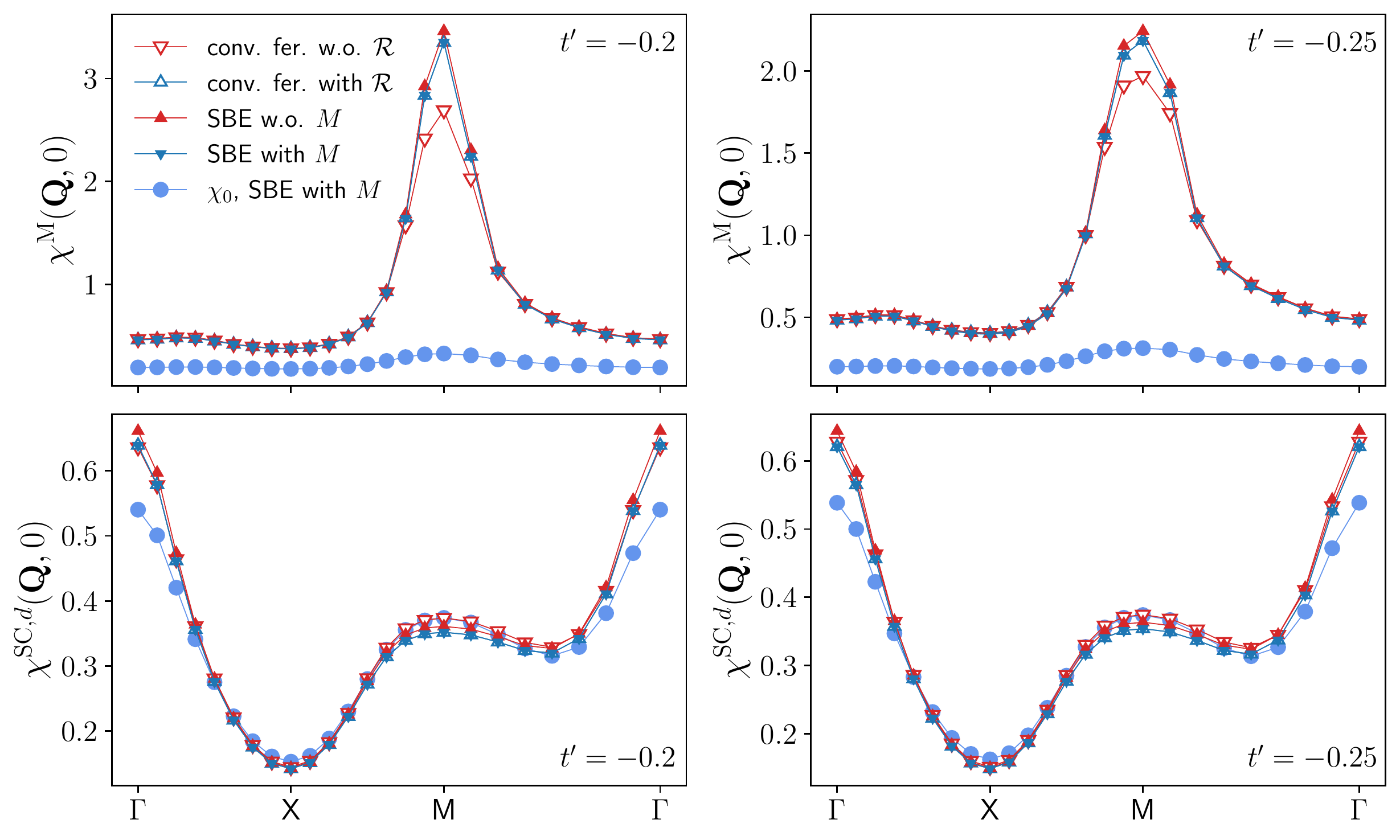}
    \caption{Magnetic $\chi^\mathrm{M}(\bQ,0)$ and $d$-wave superconducting $\chi^{\mathrm{SC},d}(\bQ,0)$ static susceptibilities including both the results obtained from the SBE and the conventional fermionic fRG formulations of the fRG with and without rest function, for $U=3$ and $t'=-0.2$ (left panels) and $t'=-0.25$ (right panels), with $\mu=4t'$, and $T=0.2$. At the end of the flow, the filling equals $0.45$ and $0.44$ at $t'=-0.2$ and $t'=-0.25$ respectively, for both the SBE and conventional fermionic decompositions, with or without rest function. We note that $\chi^\mathrm{M}(\bQ,0)$ is determined from the screened interaction, whereas $\chi^{\mathrm{SC},d}(\bQ,0)$ from post-processing. The relative difference between the data with and without rest function is always below $4\%$ for the SBE decomposition. For the conventional fermionic fRG, however, this relative difference reaches $20\%$ ($10\%$) in the magnetic and $6\%$ ($6\%$) in the superconducting channel for $t'=-0.2$ ($t'=-0.25$). Results for the bare bubble contribution $\chi_{0}$ to the susceptibilities are shown as well.}
    \label{fig: chi doping U3}
\end{figure*}

We now analyse the physical susceptibilities. The results for the magnetic $\chi^\mathrm{M}$, density $\chi^\mathrm{D}$, and $s$-wave superconducting $\chi^{\mathrm{SC},s}$ static susceptibilities along the $\mathrm{\Gamma}$-X-M-$\mathrm{\Gamma}$ path in the Brillouin zone are shown in Fig.~\ref{fig: chi doping U2}, for the same parameters used for the Yukawa couplings (the corresponding results for the frequency dependent susceptibilities are provided in the Appendix~\ref{sec: app}). The AF peak in the magnetic susceptibility appears reduced with respect to half filling, while the subleading density and $s$-wave superconducting susceptibilities exhibit a rather weak momentum dependence. The agreement of the results with and without rest function is very good also at finite doping (the deviations are below $1\%$ for all temperatures considered here), justifying the application of the SBE approximation.

The corresponding $d$-wave superconducting $\chi^{\mathrm{SC},d}$ susceptibility displayed in Fig.~\ref{fig: chi doping U2 bis} can not be determined from the ($s$-wave) screened interaction as above and is obtained by post-processing\footnote{Results for  $\chi^{\mathrm{SC},d}$ from the flow can be obtained by computing the flow of the respective response vertex~\cite{Halboth00A,Tagliavini2019}.}. We note that for the $s$-wave susceptibilities, the differences to the post-processing results that arise at the one-loop level are the same as the ones observed in the conventional fermionic fRG~\cite{Tagliavini2019}. For the considered parameters at $U=2$, the $d$-wave superconducting susceptibility appears to be well described by the bare bubble contribution $\chi_0$, shown for the lowest temperature $T=0.2$. Comparing to the results from the full computation, the vertex corrections turn out to be very small (and consequently also the ones due to inclusion of the rest function), with an appreciable contribution only around the $\mathrm{\Gamma}$-point. Moreover, the absolute values of  $\chi^{\mathrm{SC},d}$ are not very large with respect to the dominating magnetic channel of Fig.~\ref{fig: chi doping U2}, but we expect the peak at $\bQ=\bs{0}$ to increase and become more pronounced for larger values of $U$ as well as at lower temperatures.

In Fig.~\ref{fig: chi doping U3} we present the results for the most relevant magnetic and $d$-wave superconducting susceptibilities as a function of momentum for $U=3$ and in addition to a next-nearest neighbor hopping of $t'=-0.2$ also for $t'=-0.25$ (at van Hove filling). In particular, we assess the importance of the rest function both within the SBE and the conventional fermionic formulation of the fRG. Comparing to the results for $U=2$ (for $t'=-0.2$), we detect an enhanced tendency towards magnetic ordering. At the same time, changing the value of the next-nearest neighbor hopping to $t'=-0.25$ induces a reduction of the AF peak. We observe also a trend towards an incommensurate peak, as expected for larger dopings \cite{Metzner12,Halboth00A,Halboth00B,Yamase16}. In contrast, the $d$-wave pairing susceptibility is almost invariant under these changes. The considered parameter regime is still far away from any instability and we expect the $d$-wave pairing susceptibility to increase only at lower temperatures. Due to the increasing computational cost, the superconducting transition temperature is currently not accessible. Including the flow of the fermionic rest functions $M^\mathrm{X}$ (for which the SBE approach is equivalent to the conventional fRG formalism based on the 1PI vertex function), leads to slightly larger corrections as for $U=2$, but still below $4\%$. In particular, the contribution of the $d$-wave superconducting channel which at $\bs{Q}=\bs{0}$ is dominated by the rest function turns out to be negligible also for $U=3$, consistently with Figs.~\ref{fig: dwave for factorization 1} and \ref{fig: dwave for factorization 2} in the Appendix~\ref{sec: app}. In contrast, neglecting the rest function in the conventional fermionic formulation induces sizable deviations in the leading channel.

The small differences between the results with and without rest function in the $d$-wave superconducting susceptibilities demonstrate that the SBE effective interactions correctly account for the vertex corrections. These can be inferred from the difference to the respective bubble contribution $\chi_0$, reaching a relative difference of $15\%$ ($13\%$) at the $\mathrm{\Gamma}$-point for $t'=-0.2$ ($t'=-0.25$). In this parameter regime, the vertex corrections originate in the $s$-wave contributions. In fact, the computation with only a $s$-wave form factor yields no visible difference in the results (not shown). A more detailed analysis may be obtained from a fluctuation diagnostics~\cite{Gunnarsson2015,Rohringer2020rev,Schaefer2021}, which however goes beyond the scope of the present work. We finally remark that in contrast to the $d$-wave superconducting susceptibility, for the magnetic susceptibility the vertex corrections dominating the physical behavior exceed by far the bubble contribution~\cite{Heinzelmann22}).

We thus conclude that in the entire weak- to intermediate coupling regime, the susceptibilities obtained without $M^\mathrm{X}$ correctly describe the physical behavior, confirming the reliability of a description in terms of the SBE effective interactions. On a practical level, this means that the momentum and frequency dependence of the two-particle vertex can be efficiently parametrized by $\nabla^{\mathrm{X}}$, i.e. through the screened interactions and Yukawa couplings (with the exception of the pseudo-critical transition, where the weak-coupling approximation breaks down and quantitatively accurate results are not feasible even by taking into account the rest function). However, in more strongly correlated regimes where the $d$-wave rest function has to be taken into account to describe pairing fluctuations, this simplification does not apply any more and requires the computation of the multiboson contributions $M^{\mathrm{SC},d}$.

To simplify the treatment of those multiboson contributions that have a vanishing SBE term, such as the $d$-wave pairing channel, one can devise an alternative strategy to \emph{bosonize} them. One can write, for example
\begin{equation}
    M^\mathrm{X}_{\nu\nu'}(Q)= \lambda^\mathrm{X}_\nu(Q) w^\mathrm{X}(Q) \lambda_{\nu'}^\mathrm{X}(Q) + \delta M^\mathrm{X}_{\nu\nu'}(Q),
\end{equation}
where the dependence on the secondary momenta $\mathbf{k}$ and $\mathbf{k}'$ has been projected onto some form factors. The Yukawa coupling and screened interaction can be defined as
\begin{subequations}
    \label{eq: rebosonize non s-wave}
    \begin{align}
        &w^\mathrm{X}(Q) = \frac{M^\mathrm{X}_{\nu_0\nu_0}(Q)\pm M^\mathrm{X}_{\nu_0,-\nu_0}(Q)}{2}, \label{eq: rebosonized w non s-wave}\\
        &\lambda^\mathrm{X}_\nu(Q) = \frac{M^\mathrm{X}_{\nu\nu_0}(Q)\pm M^\mathrm{X}_{\nu,-\nu_0}(Q)}{2 w^\mathrm{X}(Q)},
    \end{align}
\end{subequations}
where $\nu_0$ is a fixed, finite, and eventually $Q$-dependent, Matsubara frequency. By doing that, one is able to simplify the numerical treatment and identify the collective bosonic fluctuations of those channels exhibiting only multiboson terms. Some examples of these channels are $d$-wave pairing, "Pomeranchuk density fluctuations" (for which one must choose the "+" sign above), or odd-frequency pairing (for which the "$-$" sign in Eqs.~\eqref{eq: rebosonize non s-wave} holds). The technique described above can be applied to all channels that do not have a contribution stemming from the bare interaction, that is, within the Hubbard model, all but the  $s$-wave channels. We remark that the quantity on the left-hand side of Eq.~\eqref{eq: rebosonized w non s-wave} cannot be easily related to the susceptibility in its channel, differently than the SBE screened interaction where Eq.~\eqref{eq: screened interaction vs chi} applies.

\section{Conclusions}
\label{sec:concl}

We have applied the recently introduced SBE representation, which relies on a diagrammatic decomposition in contributions mediated by the exchange of a single boson in the different channels, to the fRG analysis of the 2D Hubbard model at weak coupling. Specifically, the (one-loop) flow equations for the two-particle vertex are recast into SBE contributions and a residual four-point fermion vertex. The SBE representation provides a valuable alternative to the widely-used parquet and asymptotic decompositions in the conventional fermionic fRG. In particular, the divergent behavior arising in proximity of the pseudo-critical temperature, affects only the screened interaction while the Yukawa couplings and the SBE rest functions, which depend on more frequency and momentum variables, remain finite. In contrast, in the conventional fermionic formulation of the fRG, all objects grow very large in the RG flow by approaching the pseudo-critical instability. Moreover, the characteristic picture of bosons mediating effective interactions offers not only interpretative advantages for identifying the relevant degrees of freedom, but remains valid even at strong coupling \cite{Krien2021a,Harkov2021a}.

The quality of our SBE-based approximation to the fRG flow has been tested by performing calculations with and without the inclusion of the flow for $M^\mathrm{X}$ and by comparing the obtained results. We have found that for the computation of the physical susceptibilities in the different channels, the SBE rest function can be safely neglected in the weak-coupling regime, both at half filling and finite doping. This allows to significantly reduce the computational effort and hence to devise more efficient algorithmic implementations of the fRG flow. At larger couplings and lower temperatures, where the $d$-wave pairing correlations are expected to become more relevant, the rest function should however be included.

The numerical speed-up of the fRG algorithm offered by the SBE formalism will most likely play a pivotal role in all the extensions of the fRG approach beyond the "conventional" one-loop truncation considered in our study. Indeed, the range of current and near-future foreseeable promising developments of the fRG-based approaches is quite broad. These include the multiloop extension \cite{Kugler2018_I,Kugler2018_II,Tagliavini2019,Hille2020} of the fRG, for which the corresponding flow equations have been derived in Ref.~\cite{Gievers22}, as well as the systematic inclusion of multiboson contributions. In particular, the SBE formulation paves a promising route for the multiloop extension of the combination with the dynamical mean-field theory (DMFT) \cite{Metzner1989,Georges1996} in the so-called DMF$^2$RG \cite{Taranto2014,Vilardi2019}. This would allow to more easily access the nonperturbative regime on a quantitative level by means of the DMF$^2$RG. Here, the reduced numerical effort offered by the SBE decomposition greatly facilitates both the description of the initial DMFT vertex function \cite{Rohringer2012} and the application of the DMF$^2$RG. Further, the SBE formulation offers the possibility to study mixed boson-fermion systems in presence of additional bosonic fields and the combination with recent advancements in the description of symmetry-broken phases \cite{Wang2014,Yamase16,Bonetti20,Vilardi20}.

\section*{Acknowledgments}
The authors thank C.~Hille for setting up a first version of the code, and H.~Braun, M. Gievers, A.~Kauch, F.~Krien, F. Kugler, T.~Sch\"afer, and J. von Delft for valuable discussions. We acknowledge funding from the German Research Foundation (DFG) through the Research Unit FOR 5413/1, Grant No. 465199066, through Project No. AN 815/6-1, and from the Austrian Science Fund (FWF) through Project No. I 2794-N35.

\subsection*{Author Contributions}
S. H. implemented the SBE-based fRG algorithm. K. F. performed the calculations and analysed the data. All authors contributed to the discussion of the results and the preparation of the manuscript.

\subsection*{Data Availability}
The data generated for this study can be made available on reasonable request.

\appendix

\section{Implementation of the SBE approximation in the fRG framework}
\label{sec: app2}

\begin{figure}[t]
    \centering
    \includegraphics[width = 0.48\textwidth]{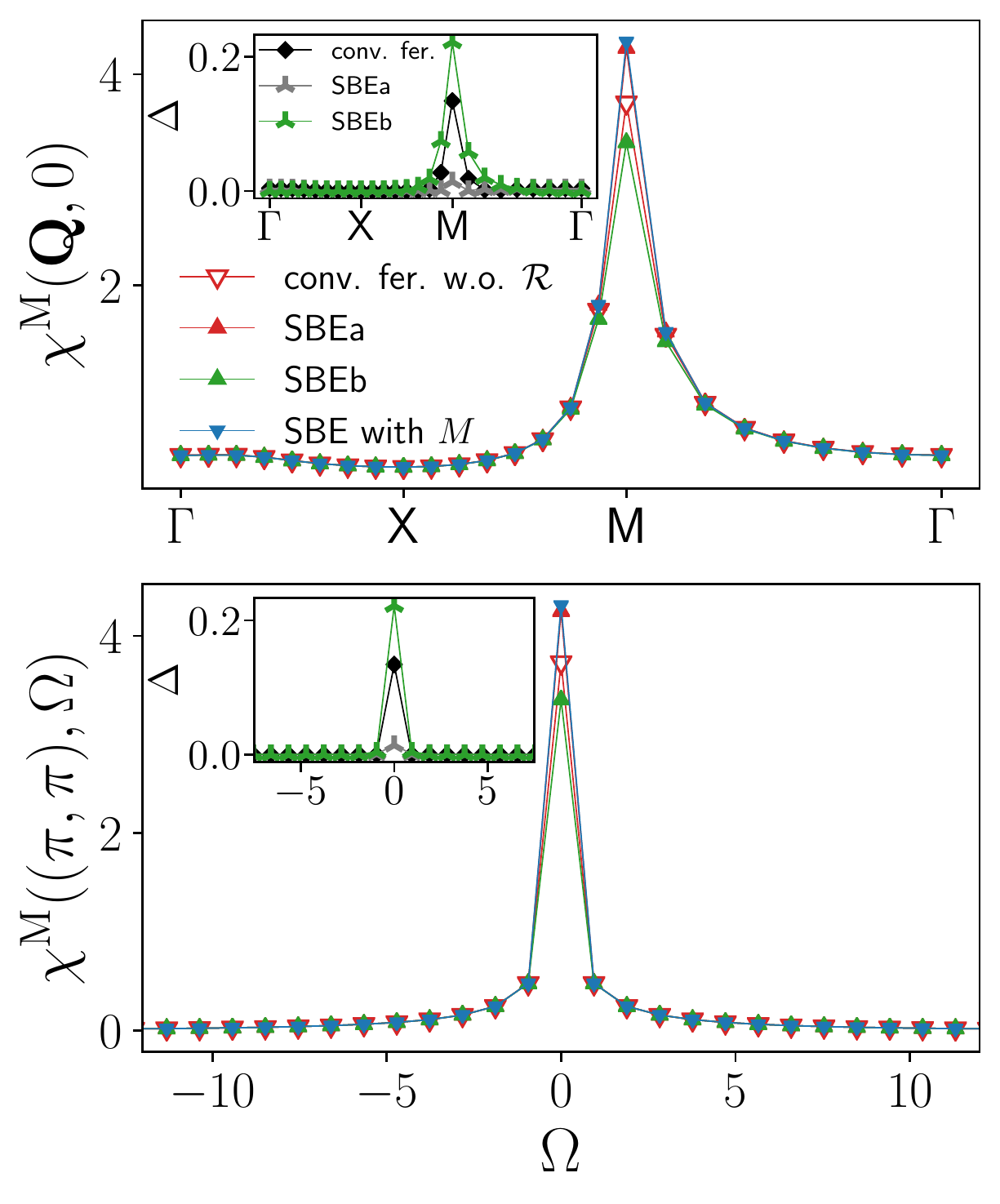}
    \caption{Bosonic momentum and frequency dependence of the magnetic susceptibility $\chi^\mathrm{M}$ as obtained from the SBE and conventional fermionic fRG formulations with and without rest function, for $U=2$ and $T=0.15$ ($t'=0$, $\mu=0$). For the SBE formulation, results are either obtained by discarding the flow of $M$ from the SBEa and SBEb approximations (see text for their definitions) or by calculating explicitly the flow of the rest function $M$. We stress also that the results labeled ``SBEa'' in the main panels are identical to those associated to the legend ``SBE w.o. $M$'' in Fig.~\ref{fig: chi momentum 2}. The insets show the relative difference $\Delta$ between results with and without rest function.}
    \label{fig: SBEavsSBEb}
\end{figure}

\begin{figure*}[t]
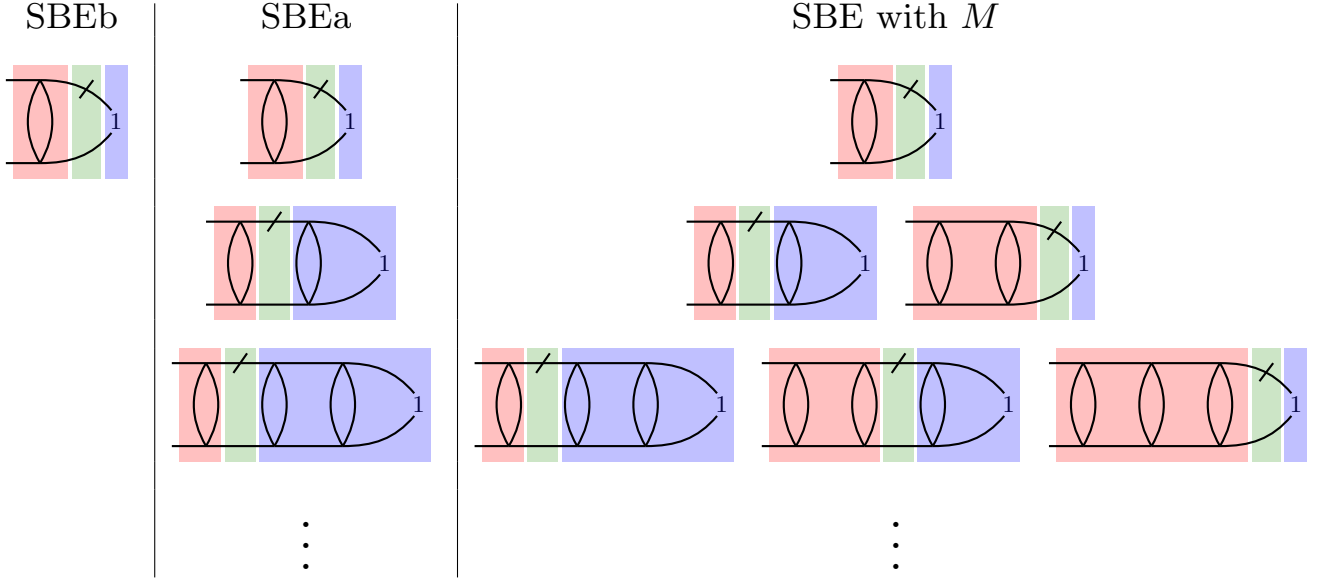

\centering
\begin{tabular}{c|c|c}
         \scalebox{1.5}{SBEb} & \scalebox{1.5}{SBEa} & \scalebox{1.5}{SBE with $M$} \\
           &  &  \\
          \tikzm{Diag1bis2}{
		    \node at (1.45,0) {$1$};
		    \fill[nearly transparent,blue] (1.45+0.15,0.75) rectangle (1.45-0.15,-0.75);
            \fill[nearly transparent,newgreen] (1.45-0.2,0.75) rectangle (1.45-0.2-0.38,-0.75);
            \fill[nearly transparent,red] (1.45-0.2-0.38-0.05,0.75) rectangle (1.45-1.35,-0.75);
		    \draw[linePlain] (1.45-0.3,0.55) -- (1.45-0.48,0.31);
		    \draw[linePlain] (1.45-0.06,0+0.15) to [out=130, in=0] (1.45-1.0,0.55);
		    \draw[linePlain] (1.45-0.06,0-0.15) to [out=-130, in=0] (1.45-1.0,-0.55);
		    \draw[linePlain] (1.45-1.0,0.55) to [out=-60, in=60] (1.45-1.0,-0.55);
		    \draw[linePlain] (1.45-1.0,0.55) to [out=240, in=-240] (1.45-1.0,-0.55);
		    \draw[linePlain] (1.45-1.0,0.55) -- (0,0.55);
		    \draw[linePlain] (1.45-1.0,-0.55) -- (0,-0.55);
		} & \tikzm{Diag1bis}{
		    \node at (1.45,0) {$1$};
		    \fill[nearly transparent,blue] (1.45+0.15,0.75) rectangle (1.45-0.15,-0.75);
            \fill[nearly transparent,newgreen] (1.45-0.2,0.75) rectangle (1.45-0.2-0.38,-0.75);
            \fill[nearly transparent,red] (1.45-0.2-0.38-0.05,0.75) rectangle (1.45-1.35,-0.75);
		    \draw[linePlain] (1.45-0.3,0.55) -- (1.45-0.48,0.31);
		    \draw[linePlain] (1.45-0.06,0+0.15) to [out=130, in=0] (1.45-1.0,0.55);
		    \draw[linePlain] (1.45-0.06,0-0.15) to [out=-130, in=0] (1.45-1.0,-0.55);
		    \draw[linePlain] (1.45-1.0,0.55) to [out=-60, in=60] (1.45-1.0,-0.55);
		    \draw[linePlain] (1.45-1.0,0.55) to [out=240, in=-240] (1.45-1.0,-0.55);
		    \draw[linePlain] (1.45-1.0,0.55) -- (0,0.55);
		    \draw[linePlain] (1.45-1.0,-0.55) -- (0,-0.55);
		} & \tikzm{Diag1}{
		    \node at (1.45,0) {$1$};
		    \fill[nearly transparent,blue] (1.45+0.15,0.75) rectangle (1.45-0.15,-0.75);
            \fill[nearly transparent,newgreen] (1.45-0.2,0.75) rectangle (1.45-0.2-0.38,-0.75);
            \fill[nearly transparent,red] (1.45-0.2-0.38-0.05,0.75) rectangle (1.45-1.35,-0.75);
		    \draw[linePlain] (1.45-0.3,0.55) -- (1.45-0.48,0.31);
		    \draw[linePlain] (1.45-0.06,0+0.15) to [out=130, in=0] (1.45-1.0,0.55);
		    \draw[linePlain] (1.45-0.06,0-0.15) to [out=-130, in=0] (1.45-1.0,-0.55);
		    \draw[linePlain] (1.45-1.0,0.55) to [out=-60, in=60] (1.45-1.0,-0.55);
		    \draw[linePlain] (1.45-1.0,0.55) to [out=240, in=-240] (1.45-1.0,-0.55);
		    \draw[linePlain] (1.45-1.0,0.55) -- (0,0.55);
		    \draw[linePlain] (1.45-1.0,-0.55) -- (0,-0.55);
		} \\
		   &  &  \\
           & \tikzm{Diag2p1bis}{
		    \node at (2.35,0) {$1$};
		    \fill[nearly transparent,blue] (2.35+0.15,0.75) rectangle (2.35-1.45+0.25,-0.75);
            \fill[nearly transparent,newgreen] (2.35-1.45+0.2,0.75) rectangle (2.35-1.45-0.2,-0.75);
            \fill[nearly transparent,red] (2.35-1.45-0.25,0.75) rectangle (2.35-2.25,-0.75);
		    \draw[linePlain] (2.35-1.36,0.68) -- (2.35-1.54,0.42);
		    \draw[linePlain] (2.35-0.06,0+0.15) to [out=130, in=0] (2.35-1.0,0.55);
		    \draw[linePlain] (2.35-0.06,0-0.15) to [out=-130, in=0] (2.35-1.0,-0.55);
		    \draw[linePlain] (2.35-1.0,0.55) to [out=-60, in=60] (2.35-1.0,-0.55);
		    \draw[linePlain] (2.35-1.0,0.55) to [out=240, in=-240] (2.35-1.0,-0.55);
		    \draw[linePlain] (2.35-1.0,0.55) -- (2.35-1.9,0.55);
		    \draw[linePlain] (2.35-1.0,-0.55) -- (2.35-1.9,-0.55);
		    \draw[linePlain] (2.35-1.9,0.55) to [out=-60, in=60] (2.35-1.9,-0.55);
		    \draw[linePlain] (2.35-1.9,0.55) to [out=240, in=-240] (2.35-1.9,-0.55);
		    \draw[linePlain] (2.35-1.9,0.55) -- (0,0.55);
		    \draw[linePlain] (2.35-1.9,-0.55) -- (0,-0.55);
		} & \tikzm{Diag2p1}{
		    \node at (2.35,0) {$1$};
		    \fill[nearly transparent,blue] (2.35+0.15,0.75) rectangle (2.35-1.45+0.25,-0.75);
            \fill[nearly transparent,newgreen] (2.35-1.45+0.2,0.75) rectangle (2.35-1.45-0.2,-0.75);
            \fill[nearly transparent,red] (2.35-1.45-0.25,0.75) rectangle (2.35-2.25,-0.75);
		    \draw[linePlain] (2.35-1.36,0.68) -- (2.35-1.54,0.42);
		    \draw[linePlain] (2.35-0.06,0+0.15) to [out=130, in=0] (2.35-1.0,0.55);
		    \draw[linePlain] (2.35-0.06,0-0.15) to [out=-130, in=0] (2.35-1.0,-0.55);
		    \draw[linePlain] (2.35-1.0,0.55) to [out=-60, in=60] (2.35-1.0,-0.55);
		    \draw[linePlain] (2.35-1.0,0.55) to [out=240, in=-240] (2.35-1.0,-0.55);
		    \draw[linePlain] (2.35-1.0,0.55) -- (2.35-1.9,0.55);
		    \draw[linePlain] (2.35-1.0,-0.55) -- (2.35-1.9,-0.55);
		    \draw[linePlain] (2.35-1.9,0.55) to [out=-60, in=60] (2.35-1.9,-0.55);
		    \draw[linePlain] (2.35-1.9,0.55) to [out=240, in=-240] (2.35-1.9,-0.55);
		    \draw[linePlain] (2.35-1.9,0.55) -- (0,0.55);
		    \draw[linePlain] (2.35-1.9,-0.55) -- (0,-0.55);
		}
		\tikzm{Diag2p2}{
		    \node at (2.35,0) {$1$};
		    \fill[nearly transparent,blue] (2.35+0.15,0.75) rectangle (2.35-0.15,-0.75);
            \fill[nearly transparent,newgreen] (2.35-0.2,0.75) rectangle (2.35-0.2-0.38,-0.75);
            \fill[nearly transparent,red] (2.35-0.2-0.38-0.05,0.75) rectangle (2.35-2.25,-0.75);
		    \draw[linePlain] (2.35-0.3,0.55) -- (2.35-0.48,0.31);
		    \draw[linePlain] (2.35-0.06,0+0.15) to [out=130, in=0] (2.35-1.0,0.55);
		    \draw[linePlain] (2.35-0.06,0-0.15) to [out=-130, in=0] (2.35-1.0,-0.55);
		    \draw[linePlain] (2.35-1.0,0.55) to [out=-60, in=60] (2.35-1.0,-0.55);
		    \draw[linePlain] (2.35-1.0,0.55) to [out=240, in=-240] (2.35-1.0,-0.55);
		    \draw[linePlain] (2.35-1.0,0.55) -- (2.35-1.9,0.55);
		    \draw[linePlain] (2.35-1.0,-0.55) -- (2.35-1.9,-0.55);
		    \draw[linePlain] (2.35-1.9,0.55) to [out=-60, in=60] (2.35-1.9,-0.55);
		    \draw[linePlain] (2.35-1.9,0.55) to [out=240, in=-240] (2.35-1.9,-0.55);
		    \draw[linePlain] (2.35-1.9,0.55) -- (0,0.55);
		    \draw[linePlain] (2.35-1.9,-0.55) -- (0,-0.55);
		} \\
	   &  &  \\
	   & \tikzm{Diag3p1bis}{
		    \node at (3.25,0) {$1$};
		    \fill[nearly transparent,blue] (3.25+0.15,0.75) rectangle (3.25-1.45+0.25-0.9,-0.75);
            \fill[nearly transparent,newgreen] (3.25-1.45-0.9+0.2,0.75) rectangle (3.25-1.45-0.9-0.2,-0.75);
            \fill[nearly transparent,red] (3.25-1.45-0.9-0.25,0.75) rectangle (3.25-3.15,-0.75);
		    \draw[linePlain] (3.25-1.36-0.9,0.68) -- (3.25-1.54-0.9,0.42);
		    \draw[linePlain] (3.25-0.06,0+0.15) to [out=130, in=0] (3.25-1.0,0.55);
		    \draw[linePlain] (3.25-0.06,0-0.15) to [out=-130, in=0] (3.25-1.0,-0.55);
		    \draw[linePlain] (3.25-1.0,0.55) to [out=-60, in=60] (3.25-1.0,-0.55);
		    \draw[linePlain] (3.25-1.0,0.55) to [out=240, in=-240] (3.25-1.0,-0.55);
		    \draw[linePlain] (3.25-1.0,0.55) -- (3.25-1.9,0.55);
		    \draw[linePlain] (3.25-1.0,-0.55) -- (3.25-1.9,-0.55);
		    \draw[linePlain] (3.25-1.9,0.55) to [out=-60, in=60] (3.25-1.9,-0.55);
		    \draw[linePlain] (3.25-1.9,0.55) to [out=240, in=-240] (3.25-1.9,-0.55);
		    \draw[linePlain] (3.25-1.9,0.55) -- (3.25-2.8,0.55);
		    \draw[linePlain] (3.25-1.9,-0.55) -- (3.25-2.8,-0.55);
		    \draw[linePlain] (3.25-2.8,0.55) to [out=-60, in=60] (3.25-2.8,-0.55);
		    \draw[linePlain] (3.25-2.8,0.55) to [out=240, in=-240] (3.25-2.8,-0.55);
		    \draw[linePlain] (3.25-2.8,0.55) -- (0,0.55);
		    \draw[linePlain] (3.25-2.8,-0.55) -- (0,-0.55);
		} & \tikzm{Diag3p1}{
		    \node at (3.25,0) {$1$};
		    \fill[nearly transparent,blue] (3.25+0.15,0.75) rectangle (3.25-1.45+0.25-0.9,-0.75);
            \fill[nearly transparent,newgreen] (3.25-1.45-0.9+0.2,0.75) rectangle (3.25-1.45-0.9-0.2,-0.75);
            \fill[nearly transparent,red] (3.25-1.45-0.9-0.25,0.75) rectangle (3.25-3.15,-0.75);
		    \draw[linePlain] (3.25-1.36-0.9,0.68) -- (3.25-1.54-0.9,0.42);
		    \draw[linePlain] (3.25-0.06,0+0.15) to [out=130, in=0] (3.25-1.0,0.55);
		    \draw[linePlain] (3.25-0.06,0-0.15) to [out=-130, in=0] (3.25-1.0,-0.55);
		    \draw[linePlain] (3.25-1.0,0.55) to [out=-60, in=60] (3.25-1.0,-0.55);
		    \draw[linePlain] (3.25-1.0,0.55) to [out=240, in=-240] (3.25-1.0,-0.55);
		    \draw[linePlain] (3.25-1.0,0.55) -- (3.25-1.9,0.55);
		    \draw[linePlain] (3.25-1.0,-0.55) -- (3.25-1.9,-0.55);
		    \draw[linePlain] (3.25-1.9,0.55) to [out=-60, in=60] (3.25-1.9,-0.55);
		    \draw[linePlain] (3.25-1.9,0.55) to [out=240, in=-240] (3.25-1.9,-0.55);
		    \draw[linePlain] (3.25-1.9,0.55) -- (3.25-2.8,0.55);
		    \draw[linePlain] (3.25-1.9,-0.55) -- (3.25-2.8,-0.55);
		    \draw[linePlain] (3.25-2.8,0.55) to [out=-60, in=60] (3.25-2.8,-0.55);
		    \draw[linePlain] (3.25-2.8,0.55) to [out=240, in=-240] (3.25-2.8,-0.55);
		    \draw[linePlain] (3.25-2.8,0.55) -- (0,0.55);
		    \draw[linePlain] (3.25-2.8,-0.55) -- (0,-0.55);
		}
		\tikzm{Diag3p2}{
		    \node at (3.25,0) {$1$};
		    \fill[nearly transparent,blue] (3.25+0.15,0.75) rectangle (3.25-1.45+0.25,-0.75);
            \fill[nearly transparent,newgreen] (3.25-1.45+0.2,0.75) rectangle (3.25-1.45-0.2,-0.75);
            \fill[nearly transparent,red] (3.25-1.45-0.25,0.75) rectangle (3.25-3.15,-0.75);
		    \draw[linePlain] (3.25-1.36,0.68) -- (3.25-1.54,0.42);
		    \draw[linePlain] (3.25-0.06,0+0.15) to [out=130, in=0] (3.25-1.0,0.55);
		    \draw[linePlain] (3.25-0.06,0-0.15) to [out=-130, in=0] (3.25-1.0,-0.55);
		    \draw[linePlain] (3.25-1.0,0.55) to [out=-60, in=60] (3.25-1.0,-0.55);
		    \draw[linePlain] (3.25-1.0,0.55) to [out=240, in=-240] (3.25-1.0,-0.55);
		    \draw[linePlain] (3.25-1.0,0.55) -- (3.25-1.9,0.55);
		    \draw[linePlain] (3.25-1.0,-0.55) -- (3.25-1.9,-0.55);
		    \draw[linePlain] (3.25-1.9,0.55) to [out=-60, in=60] (3.25-1.9,-0.55);
		    \draw[linePlain] (3.25-1.9,0.55) to [out=240, in=-240] (3.25-1.9,-0.55);
		    \draw[linePlain] (3.25-1.9,0.55) -- (3.25-2.8,0.55);
		    \draw[linePlain] (3.25-1.9,-0.55) -- (3.25-2.8,-0.55);
		    \draw[linePlain] (3.25-2.8,0.55) to [out=-60, in=60] (3.25-2.8,-0.55);
		    \draw[linePlain] (3.25-2.8,0.55) to [out=240, in=-240] (3.25-2.8,-0.55);
		    \draw[linePlain] (3.25-2.8,0.55) -- (0,0.55);
		    \draw[linePlain] (3.25-2.8,-0.55) -- (0,-0.55);
		}
		\tikzm{Diag3p3}{
		    \node at (3.25,0) {$1$};
		    \fill[nearly transparent,blue] (3.25+0.15,0.75) rectangle (3.25-0.15,-0.75);
            \fill[nearly transparent,newgreen] (3.25-0.2,0.75) rectangle (3.25-0.2-0.38,-0.75);
            \fill[nearly transparent,red] (3.25-0.2-0.38-0.05,0.75) rectangle (3.25-3.15,-0.75);
		    \draw[linePlain] (3.25-0.3,0.55) -- (3.25-0.48,0.31);
		    \draw[linePlain] (3.25-0.06,0+0.15) to [out=130, in=0] (3.25-1.0,0.55);
		    \draw[linePlain] (3.25-0.06,0-0.15) to [out=-130, in=0] (3.25-1.0,-0.55);
		    \draw[linePlain] (3.25-1.0,0.55) to [out=-60, in=60] (3.25-1.0,-0.55);
		    \draw[linePlain] (3.25-1.0,0.55) to [out=240, in=-240] (3.25-1.0,-0.55);
		    \draw[linePlain] (3.25-1.0,0.55) -- (3.25-1.9,0.55);
		    \draw[linePlain] (3.25-1.0,-0.55) -- (3.25-1.9,-0.55);
		    \draw[linePlain] (3.25-1.9,0.55) to [out=-60, in=60] (3.25-1.9,-0.55);
		    \draw[linePlain] (3.25-1.9,0.55) to [out=240, in=-240] (3.25-1.9,-0.55);
		    \draw[linePlain] (3.25-1.9,0.55) -- (3.25-2.8,0.55);
		    \draw[linePlain] (3.25-1.9,-0.55) -- (3.25-2.8,-0.55);
		    \draw[linePlain] (3.25-2.8,0.55) to [out=-60, in=60] (3.25-2.8,-0.55);
		    \draw[linePlain] (3.25-2.8,0.55) to [out=240, in=-240] (3.25-2.8,-0.55);
		    \draw[linePlain] (3.25-2.8,0.55) -- (0,0.55);
		    \draw[linePlain] (3.25-2.8,-0.55) -- (0,-0.55);
		} \\
	    &  &  \\
	    & \scalebox{2.0}{$\vdots$} & \scalebox{2.0}{$\vdots$} \\
    \end{tabular}
    \caption{Comparison of the derivatives included during the flow of $\partial_\Lambda \lambda^\mathrm{X}_k(Q)$ via Eq.~\eqref{eq: flow eq h general} vs \eqref{eq: flow eq h general sbeapprox}. The left red box represents $\mathcal{I}^\mathrm{X}_{kp}(Q)$, the central green $\widetilde{\partial}_\Lambda \Pi^\mathrm{X}_p(Q)$, and the right blue $\lambda^\mathrm{X}_k(Q)$ (which might reduce to $1$ depending on the approximation). The diagrams in the first row are generated during the first step of the solver, those of the second and third are added in the second and third steps respectively. While the replacement of $\lambda^\mathrm{X}_k(Q)$ by $1$ in the SBEb approximation limits the derivative to the eye diagrams, all diagrams are constructed during the flow of the SBEa albeit with derivatives less complete than for the SBE including $M$.}
    \label{fig: SBEavsSBEb diagrams}
\end{figure*}

We here discuss an alternative approximation based on completely neglecting the rest function together with its implications on the flow of the screened interactions and Yukawa couplings: Instead of deriving the flow equations and then setting the flow of the rest function to zero, as in our analysis, in Ref.~\cite{Gievers22} the authors suggest neglecting the rest function $prior$ to the derivation of flow equations. In the following, these approximations will be referred to as SBE approximation after (SBEa) and before (SBEb) taking the derivative respectively. The resulting flow equations for SBEb are structurally simpler than the corresponding Eqs.~\eqref{eq: flow eq general}:
\begin{subequations}
    \label{eq: flow eq general 1}
    \begin{align}
        \partial_\Lambda w^\mathrm{X}(Q)& = \left[w^\mathrm{X}(Q)\right]^2\!\!\int_p (2\lambda^\mathrm{X}_p(Q)-1) \left[\widetilde{\partial}_\Lambda \Pi^\mathrm{X}_p(Q) \right] 1, \label{eq: flow eq D general sbeapprox}\\
        \partial_\Lambda \lambda^\mathrm{X}_k(Q)& = \int_p \mathcal{I}^\mathrm{X}_{kp}(Q) \left[\widetilde{\partial}_\Lambda \Pi^\mathrm{X}_p(Q) \right] 1,
        \label{eq: flow eq h general sbeapprox}
    \end{align}
\end{subequations}
 where we exploited $\lambda=\Bar{\lambda}$ to obtain
\begin{align}
&&\lambda^\mathrm{X}_p(Q)\left[\widetilde{\partial}_\Lambda \Pi^\mathrm{X}_p(Q) \right] 1 + 1 \left[\widetilde{\partial}_\Lambda \Pi^\mathrm{X}_p(Q) \right] \Bar{\lambda}^\mathrm{X}_p(Q) \quad\nonumber\\
 &&-1\left[\widetilde{\partial}_\Lambda \Pi^\mathrm{X}_p(Q) \right]1 = (2\lambda^\mathrm{X}_p(Q)-1) \left[\widetilde{\partial}_\Lambda \Pi^\mathrm{X}_p(Q) \right] 1, 
\end{align}
and the factor $1$ has to be replaced by the $s$-wave projection of the bubble in a form factor expansion (see Appendix~\ref{sec: flow eq in form factors}). 
In fact, the SBEb appears to be a more severe approximation, as can be seen in Fig.~\ref{fig: SBEavsSBEb}. 
The results for the SBEa and SBEb approximations are compared through their relative difference with respect to the corresponding data obtained without neglecting the rest function $M$. 
While the SBEa approximation provides quantitatively correct results above the pseudo-critical temperature in the weak coupling regime, the same cannot be concluded for the SBEb which underestimates the AF peak.

Both the SBEa and SBEb approximations are  based on the assumption that the relevant physics is well captured by the exchange of a single boson, and that the multiboson processes of the rest function can therefore be neglected. The observed discrepancy is due to the omission of all rest function like contributions in the Yukawa couplings and bosonic propagators in the self-consistent SBEb. Here the Yukawa coupling contributions are restricted to the eye diagrams as $\mathcal{I}^\mathrm{X}$ does not contain any diagrams of the rest function in the channel $\mathrm{X}$. In contrast, the SBEa flow equations for the Yukawa couplings generate all diagrams at least partially (although with less complete derivatives as when the rest function is included). We note that $\mathcal{I}^\mathrm{X}$ is not structurally different in SBEa and SBEb but, as $\lambda$ appears on the right-hand side of the flow equation, rest function like contributions are generated, see the lowest order diagrams shown in Fig.~\ref{fig: SBEavsSBEb diagrams}. It turns out that these diagrams are relevant to capture the physical behavior and that their partial treatment in the SBEa is sufficient for the considered parameter regimes. This is also true at stronger couplings, where the nonlocal corrections to the rest functions have been shown to play a negligible role~\cite{Bonetti2022}.

\section{Flow equations in form factor notation}
\label{sec: flow eq in form factors}
We here explicitly report the flow Eqs.~\eqref{eq: flow eq general} in the form factor expansion. They are given by
\begin{subequations}
    \begin{align}
        \partial_\Lambda w^\mathrm{X}(\bs{Q},i\Omega) = \left[w^\mathrm{X}(\bs{Q},i\Omega)\right]^2\!\!\sum_{i\nu,m,m'} \lambda^{\mathrm{X},m}(\bs{Q},i\Omega,i\nu) \nonumber\\
        \times\left[\widetilde{\partial}_\Lambda \Pi^{\mathrm{X},mm'}(\bs{Q},i\Omega,i\nu) \right] \lambda^{\mathrm{X},m'}(\bs{Q},i\Omega,i\nu), 
        \label{eq: flow eq w form factor sbea approx}\\
        \partial_\Lambda \lambda^{\mathrm{X},n}(\bs{Q},i\Omega,i\nu) = \sum_{i\nu',m,m'} \mathcal{I}^{\mathrm{X},n,m}(\bs{Q},i\Omega,i\nu,i\nu') \nonumber\\
        \times\left[\widetilde{\partial}_\Lambda \Pi^{\mathrm{X},mm'}(\bs{Q},i\Omega,i\nu') \right] 
        \lambda^{\mathrm{X},m'}(\bs{Q},i\Omega,i\nu'),
        \label{eq: flow eq lambda form factor sbea approx}
        \\
        \partial_\Lambda M^{\mathrm{X},nn'}(\bs{Q},i\Omega,i\nu,i\nu') = \sum_{i\nu'',m,m'} \mathcal{I}^{\mathrm{X},n,m}(\bs{Q},i\Omega,i\nu,i\nu'') \nonumber\\
        \times\left[\widetilde{\partial}_\Lambda \Pi^{\mathrm{X},mm'}(\bs{Q},i\Omega,i\nu'') \right] 
        \mathcal{I}^{\mathrm{X},m',n'}(\bs{Q},i\Omega,i\nu'',i\nu').
        \label{eq: flow eq M form factor sbea approx}
    \end{align}
\end{subequations}

For the SBEb approximation (see Appendix~\ref{sec: app2}), the right $\lambda^{\mathrm{X},n}(\bs{Q},i\Omega,\nu)$ is replaced by $\delta_{n,s}$ yielding
\begin{subequations}
    \begin{align}
        \partial_\Lambda w^\mathrm{X}(\bs{Q},i\Omega) = &\left[w^\mathrm{X}(\bs{Q},i\Omega)\right]^2 \nonumber\\
        &\times\sum_{i\nu,m} (2\lambda^{\mathrm{X},m}(\bs{Q},i\Omega,i\nu)-\delta_{m,s}) \nonumber\\
        &\times\left[\widetilde{\partial}_\Lambda \Pi^{\mathrm{X},ms}(\bs{Q},i\Omega,i\nu) \right] , 
        \label{eq: flow eq w form factor sbeb approx}\\
        \partial_\Lambda \lambda^{\mathrm{X},n}(\bs{Q},i\Omega,i\nu) = &\sum_{i\nu',m} \mathcal{I}^{\mathrm{X},n,m}(\bs{Q},i\Omega,i\nu,i\nu') \nonumber\\
        &\times\left[\widetilde{\partial}_\Lambda \Pi^{\mathrm{X},ms}(\bs{Q},i\Omega,i\nu') \right],
        \label{eq: flow eq lambda form factor sbeb approx}
    \end{align}
\end{subequations}
corresponding to Eqs.~\eqref{eq: flow eq general 1}.

\section{Additional results at half filling and finite doping}
\label{sec: app}

\subsection{Susceptibilities}

\begin{figure}[t]
    \centering
    \includegraphics[width = 0.48\textwidth]{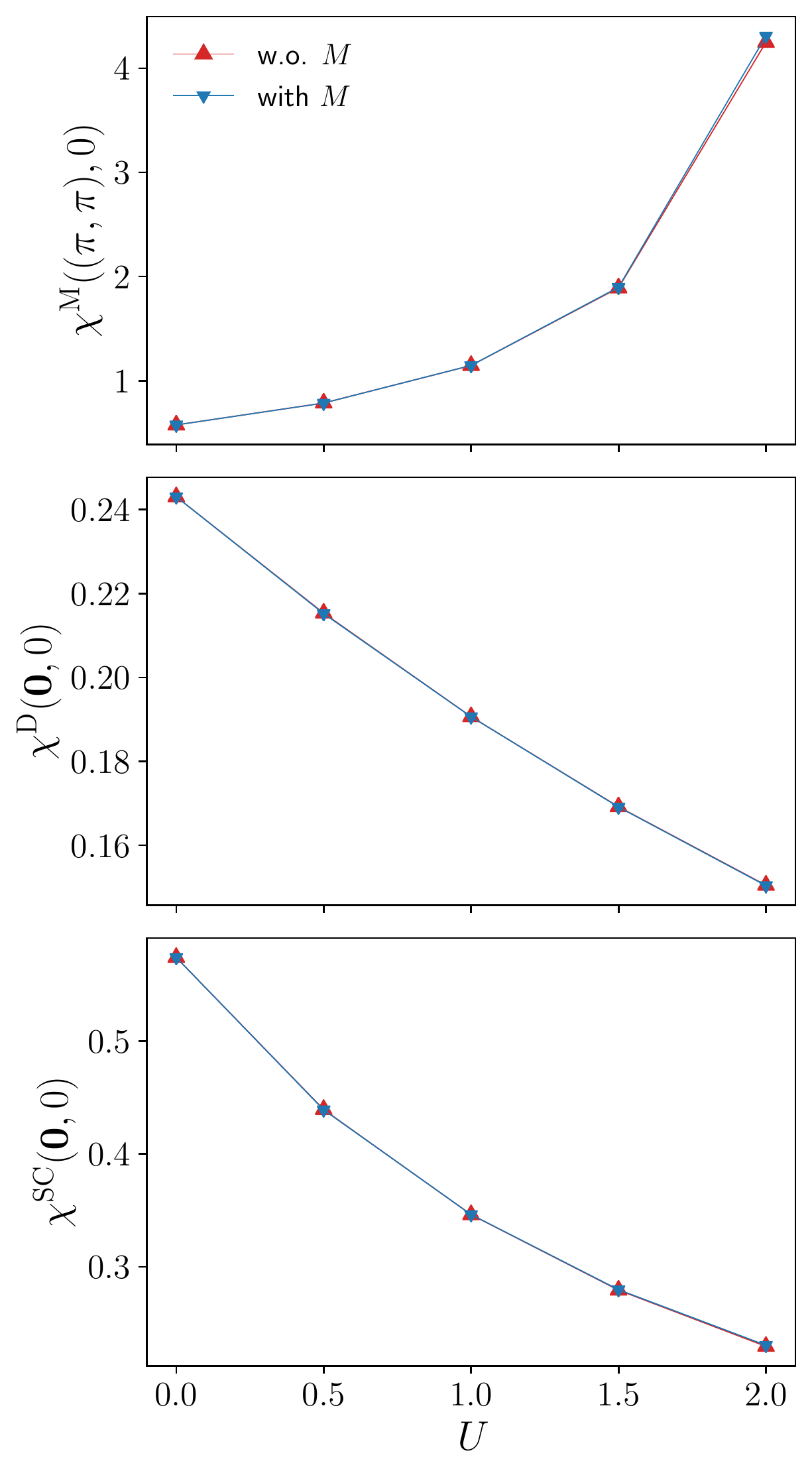}
    \caption{Magnetic $\chi^\mathrm{M}((\pi,\pi),0)$, density $\chi^\mathrm{D}((0,0),0)$, and $s$-wave superconducting $\chi^\mathrm{SC}((0,0),0)$ static susceptibilities as obtained from the SBE formulation of the fRG with and without rest function as a function of $U$, for $T=0.15$ ($t'=0$, $\mu=0$). The relative difference between the results with and without rest function is below $0.1\%$ in the density and $0.5\%$ in the superconducting channel, while it reaches $1.4 \%$ in the magnetic one (for $U=2$).}
    \label{fig: chi U}
\end{figure}

\begin{figure}[t]
    \centering
    \includegraphics[width = 0.48\textwidth]{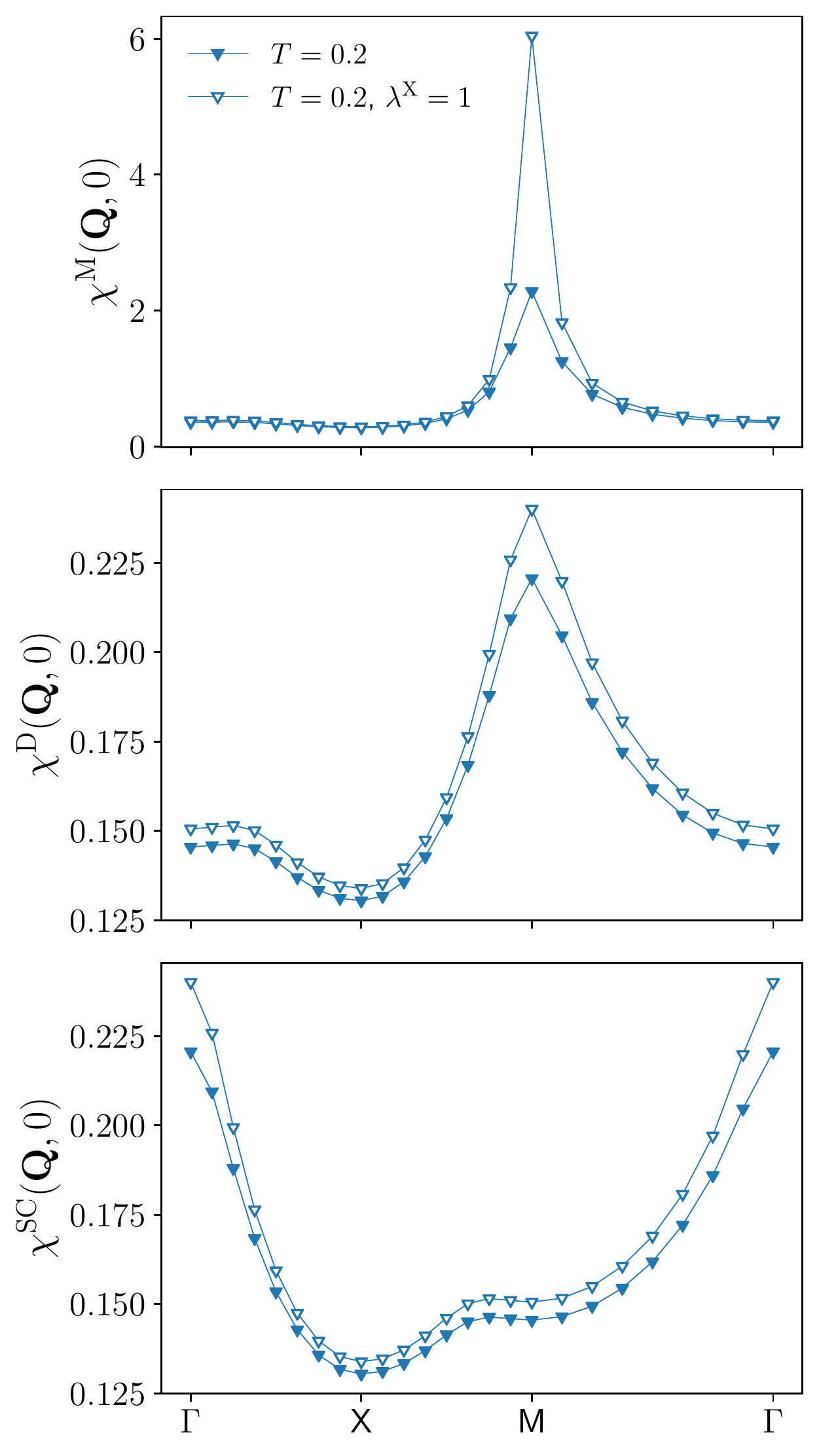}
    \caption{Magnetic $\chi^\mathrm{M}(\bQ,0)$, density $\chi^\mathrm{D}(\bQ,0)$, and $s$-wave superconducting $\chi^{\mathrm{SC}}(\bQ,0)$ static susceptibilities as obtained from the SBE formulation of the fRG (with rest functions), where the flow of the Yukawa couplings has been discarded, i.e. by fixing $\lambda^\mathrm{X} = 1$ throughout the flow, for $U=2$ and $T=0.2$ ($t'=0$, $\mu=0$), see also Fig.~\ref{fig: chi momentum} for comparison. The relative difference between the results obtained with and without imposing this condition reaches $9 \%$ in both the density and superconducting channels, and $165 \%$ in the magnetic one.}
    \label{fig: chi momentum 3}
\end{figure}

\begin{figure}[t]
    \centering
    \includegraphics[width = 0.48\textwidth]{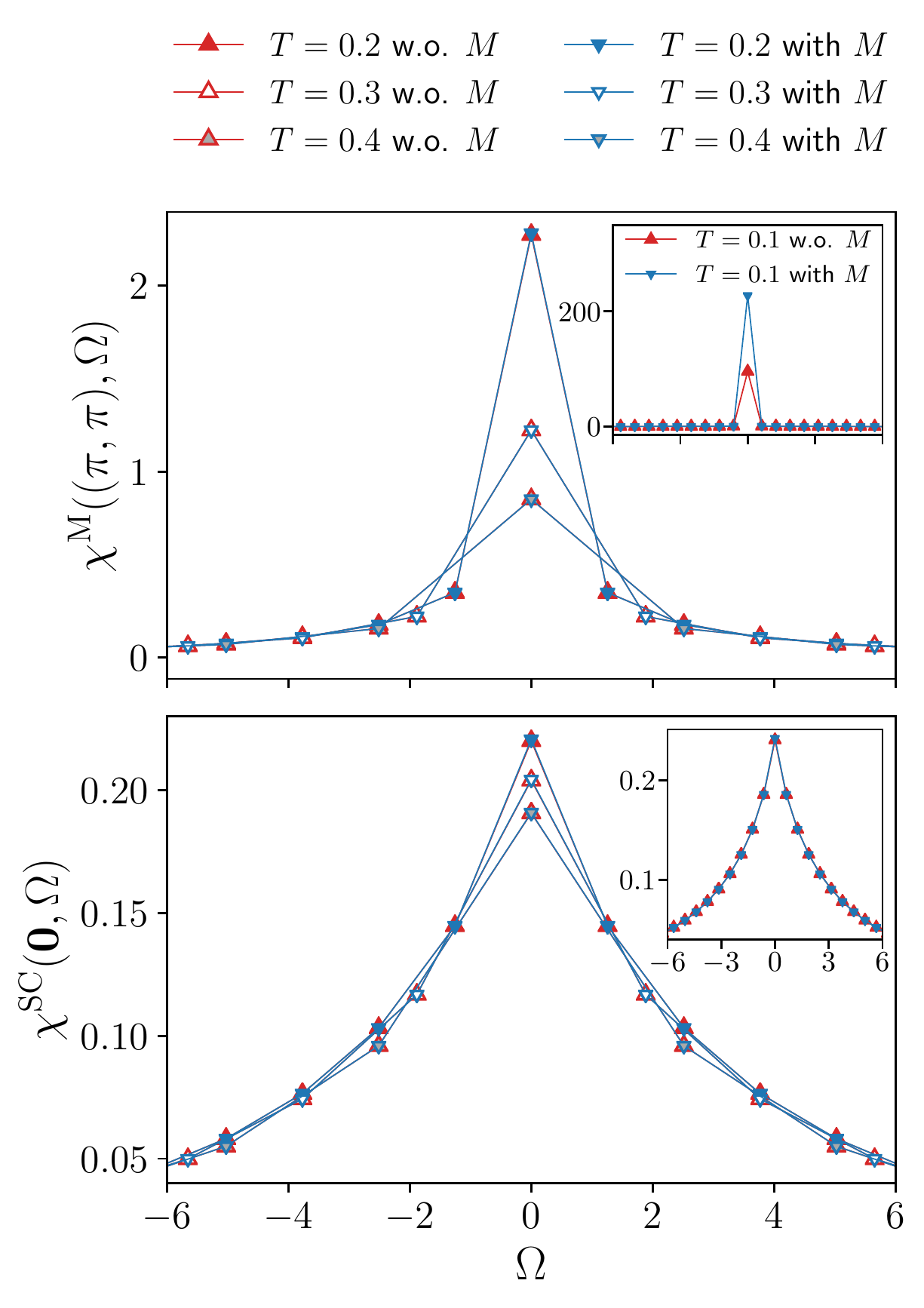}
    \caption{Bosonic frequency dependence of the magnetic $\chi^\mathrm{M}$ and superconducting $\chi^\mathrm{SC}$ susceptibilities for the same parameters as in Fig.~\ref{fig: yukawas half filling}. The relative difference $\Delta$ between the results with and without rest function does not exceed $0.8\%$, except for the magnetic channel at $T=0.1$ where $\Delta$ reaches $58\%$.}
    \label{fig: chivsOm HalfFilling}
\end{figure}

\begin{figure}[t]
    \centering
    \includegraphics[width = 0.48\textwidth]{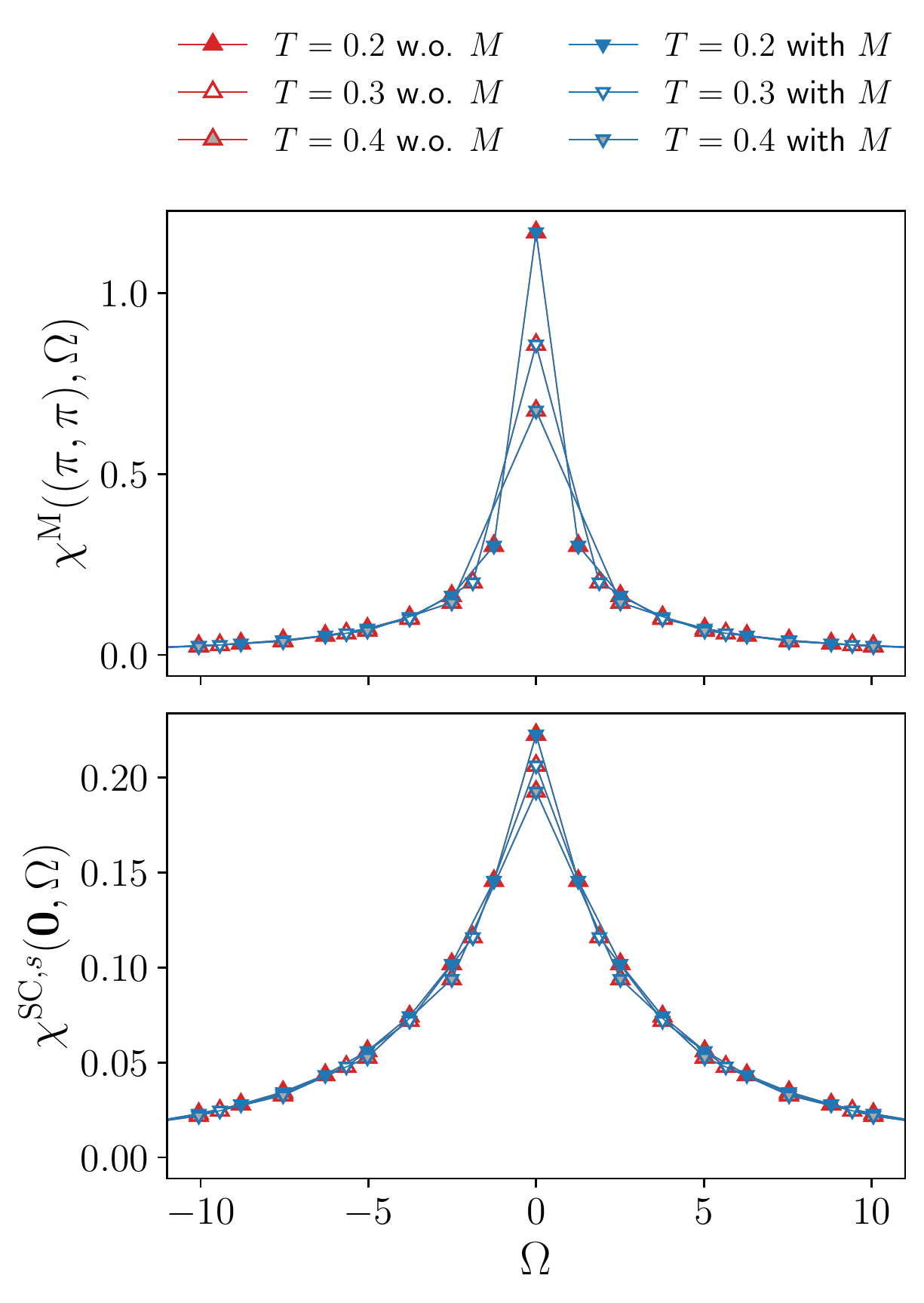}
    \caption{Bosonic frequency dependence of the magnetic $\chi^\mathrm{M}$ and $s$-wave superconducting $\chi^{\mathrm{SC},s}$ susceptibilities for the same parameters as in Fig.~\ref{fig: yukawas doping}. The relative difference between the results with and without rest function is always below $4\%$ in both the magnetic and $s$-wave superconducting channels.}
    \label{fig: chivsOm FiniteDoping}
\end{figure}

In Fig.~\ref{fig: chi U} we show the evolution of the magnetic, density, and $s$-wave superconducting susceptibilities with $U$, in analogy to Fig.~\ref{fig: chi T} for the temperature. The magnetic susceptibility is strongly enhanced with the interaction, due to the increasing AF fluctuations. In contrast, in the subleading channels the correlation effects are weaker. We observe a slight suppression of the uniform density and $s$-wave superconducting static susceptibilities induced by the interplay with their complementary channels. Also here, for the description of the physical susceptibilities the rest function can be safely neglected.

We report also the effects of the approximation $\lambda^\mathrm{X}=1$, i.e. of neglecting the flow of the Yukawa couplings. Figure~\ref{fig: chi momentum 3} displays the obtained results for the (static) susceptibilities in the different channels, for the same parameters as in Fig.~\ref{fig: chi momentum}. We observe that this additional approximation, somewhat similar to the one made in the TRILEX approach~\cite{Ayral2015} and GW~\cite{Aryasetiawan1998,Hedin1965}, leads to considerable deviations in the considered parameter regime~\cite{Schaefer2021a}: the assumption $\lambda^\mathrm{X}=1$ leads to overestimate the value of the Yukawa couplings, in particular in correspondence of the maxima, see Fig.~\ref{fig: yukawas half filling}, which is then reflected in an enhancement of the corresponding susceptibilities. In particular, since imposing $\lambda^\mathrm{X}=1$ overestimates the AF susceptibility by $165 \%$,  discarding the flow of the Yukawa couplings does not represent a viable approximation for the problem under investigation.

We finally report the (bosonic) frequency dependence in Figs.~\ref{fig: chivsOm HalfFilling} and~\ref{fig: chivsOm FiniteDoping}, corresponding to the Yukawa couplings of Figs.~\ref{fig: yukawas freq half filling} and \ref{fig: yukawas freq doping} respectively. The density channel is omitted in these two figures since $\chi^{\mathrm{D}}(\bs{Q},\Omega)$ vanishes at finite frequency $\Omega$ at $\bs{Q}=\bs{0}$, consistently \cite{Watzenboeck2022} with the conservation of the total charge of the system.

\subsection{Self-energy}

\begin{figure}[t]
    \centering
    \includegraphics[width = .48\textwidth]{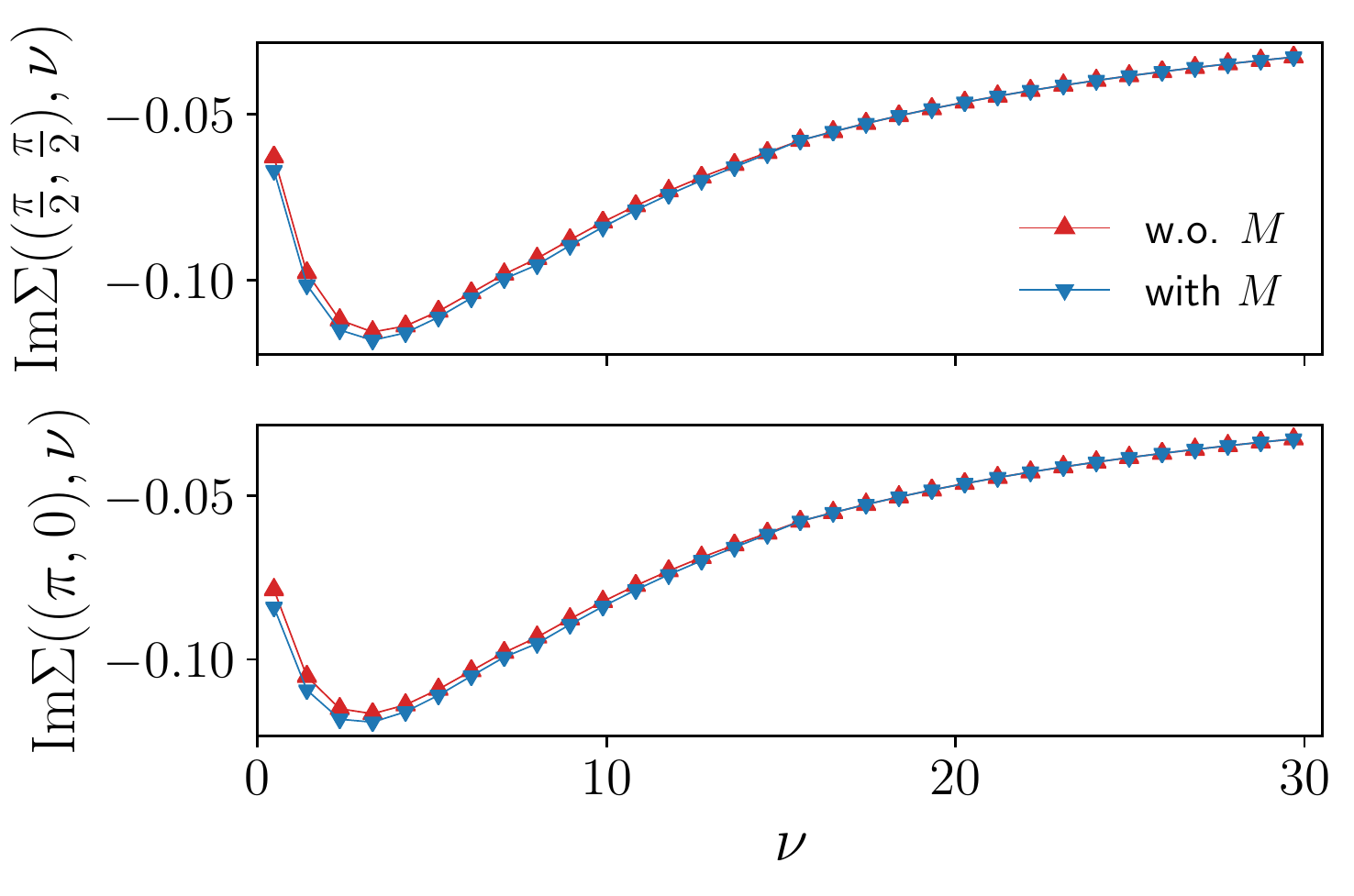}
    \caption{Imaginary part of the self-energy $\Sigma({\bs{k}},\nu)$ at the node ${\bs{k}}=(\pi/2,\pi/2)$ and antinode ${\bs{k}}=(\pi,0)$ for $U=2$ and $T=0.15$ ($t'=0$, $\mu=0$).}
    \label{fig: Im Sigma omega}
\end{figure}

The results for the self-energy, as obtained from the SBE formulation, are displayed in Fig.~\ref{fig: Im Sigma omega} for the representative parameters of $U=2$ and $T=0.15$ (at half filling). The frequency dependence of the imaginary part presents a typical Fermi-liquid behavior at small frequencies, both at the node ${\bf k}=(\pi/2,\pi/2)$ in the upper and at the antinode ${\bf k}=(\pi,0)$ in the lower panel. For these parameters, the resulting self-energy does not develop a momentum-selective gap. The computation without rest functions $M^\mathrm{X}$ almost perfectly reproduces the full one including the rest function. The deviations of the order of a few per cent are maximal for the lowest Matsubara frequency $\Omega=\pi T$ and decrease for larger frequencies. On a quantitative level, in the full Green's function $G({\bf k},i\nu)$ the differences between the computation with and without rest function are almost negligible due to the large $i\nu$ contribution of the bare Green's function.

\subsection{Rest functions}

\begin{figure*}[t]
    \centering
    \includegraphics[width = .9\textwidth]{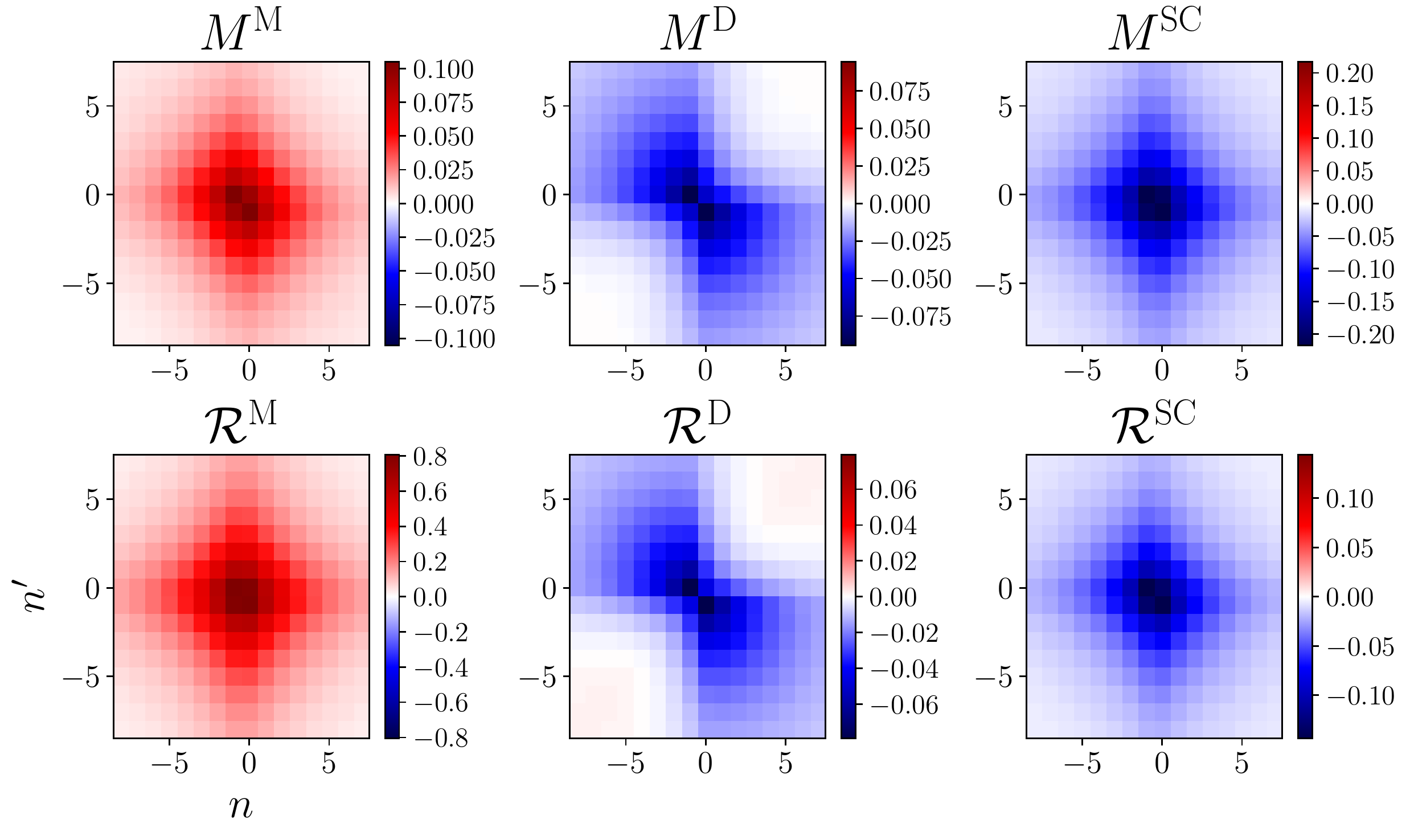}
    \caption{Rest functions for the magnetic $M^\mathrm{M}_{\nu\nu'}((\pi,\pi),0)$, density $M^\mathrm{D}_{\nu\nu'}((0,0),0)$, and $s$-wave superconducting $M^\mathrm{SC}_{\nu\nu'}((0,0),0)$ channels as obtained from the SBE (upper panels) and the conventional fermionic (lower panels) fRG formulations, with $n$ and $n^{\prime}$ labeling the fermionic Matsubara frequencies according to $\nu^{(\prime)}=(2n^{(\prime)}+1)\pi T$, for $U=2$ and $T=0.15$ ($t'=0$, $\mu=0$). Note the different orders of magnitude in the $z$-axis for the magnetic channel.}
    \label{fig: rest functions}
\end{figure*}

\begin{figure*}[b]
    \centering
   \includegraphics[width = .9\textwidth]{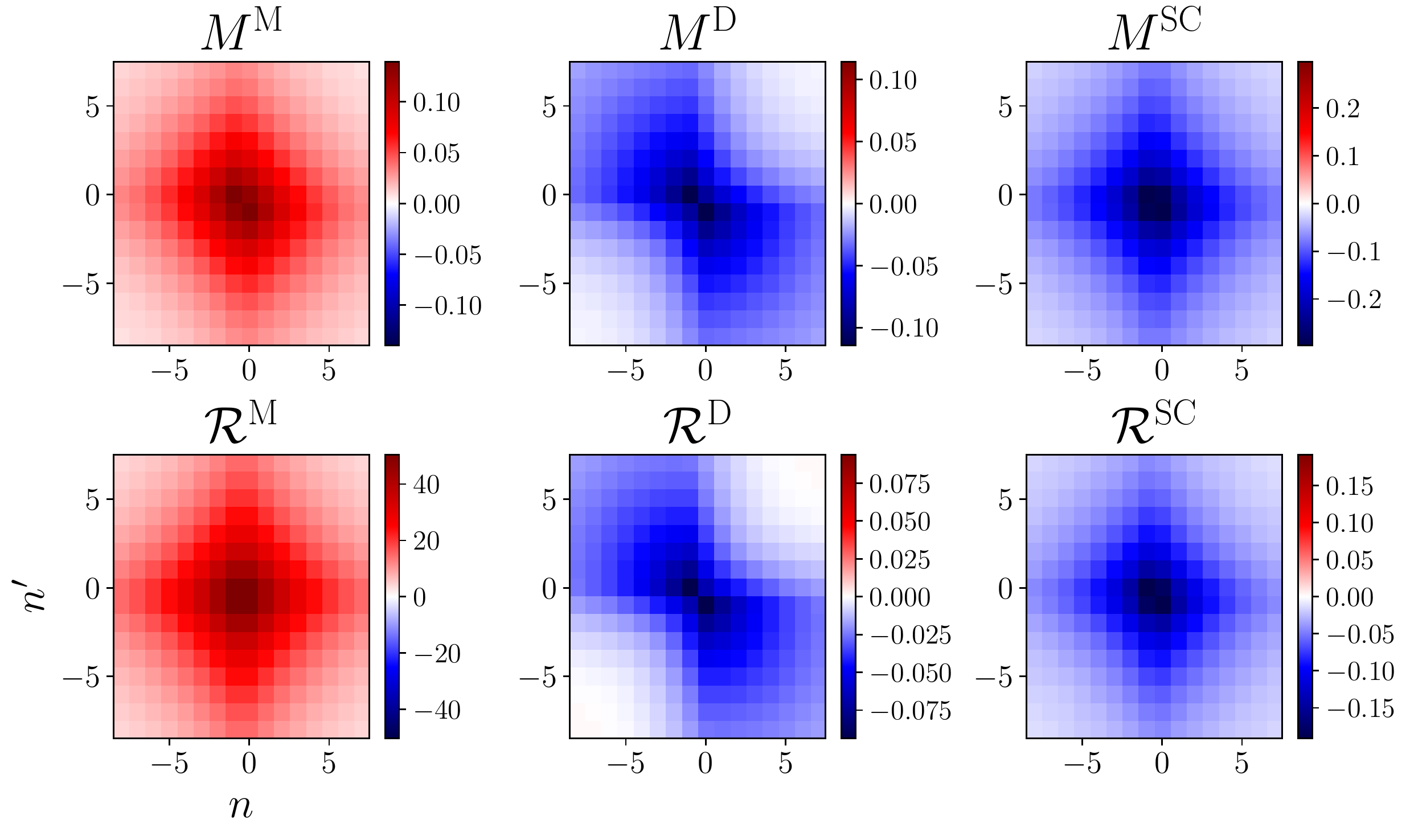}
    \caption{Same as Fig.~\ref{fig: rest functions} with $T=0.1$ instead.}
    \label{fig: rest functions 2}
\end{figure*}

In Figs.~\ref{fig: rest functions} and \ref{fig: rest functions 2} we show the rest functions in all channels, for $T=0.15$ as well as for $T=0.1$ (at half filling). While the absolute values for the density and superconducting channels are similar for the SBE and the conventional fermionic decomposition, the magnetic channel differs by an order of magnitude for $T=0.15$ (see Fig.~\ref{fig: rest functions}). In proximity of the pseudo-critical transition, this effect is even more pronounced for $T=0.1$ (see Fig.~\ref{fig: rest functions 2}). The magnetic channel becomes very large, whereas the other channels are only slightly enhanced with respect to the results for $T=0.1$.

We note that the frequency range, over which the rest function in the dominant magnetic channel extends, appears to be slightly smaller for the SBE than for the conventional fermionic representation. We expect this effect to be more pronounced for larger values of the interaction~\cite{Bonetti2022}. There, the SBE-based formulation leads to a substantial reduction of the numerical effort, since the corresponding rest function is significantly localized in frequency space. This allows one to significantly restrict the total number of frequencies taken into account in the fRG flow, facilitating the applicability of the fRG and DMF$^2$RG to the most interesting regime of intermediate to strong correlations and/or low temperatures~\cite{Bonetti2022}.

\begin{figure*}[t]
    \centering
    \includegraphics[width=1.0\textwidth]{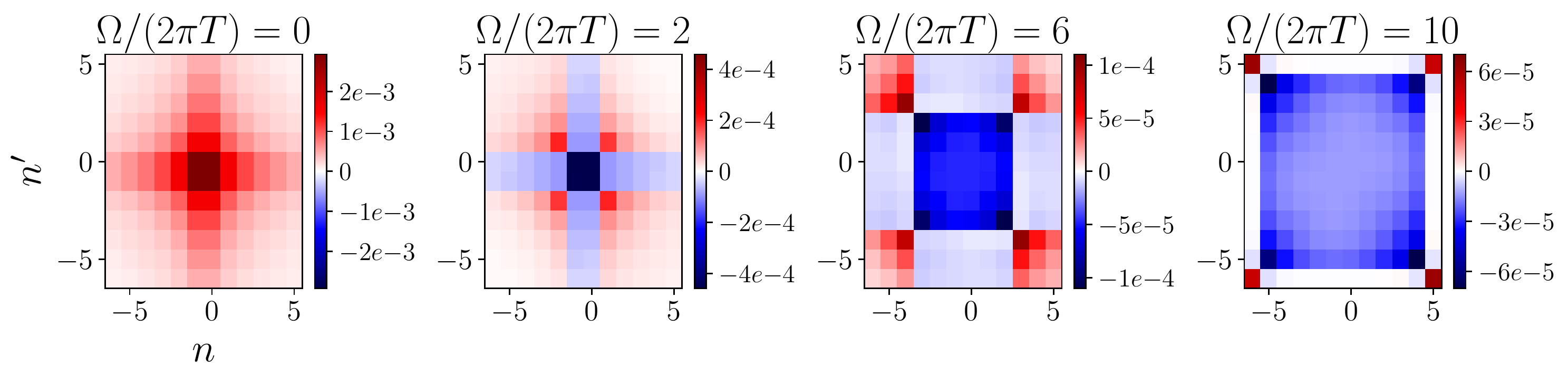}
    \caption{Rest function $M^{\mathrm{SC},d}_{\nu\nu'}((0,0),\Omega)=\phi^{\mathrm{SC},d}_{\nu\nu'}((0,0),\Omega)$ in the $d$-wave pairing channel, with $n$ and $n^{\prime}$ labeling the fermionic Matsubara frequencies according to $\nu^{(\prime)}=(2n^{(\prime)}+1)\pi T$, for $U=2$, $t^\prime=-0.2$, $\mu=4t'$, and different choices of the bosonic frequency $\Omega$.}
    \label{fig: dwave for factorization 1}
\end{figure*}

\begin{figure*}[t]
    \centering
    \includegraphics[width=1.0\textwidth]{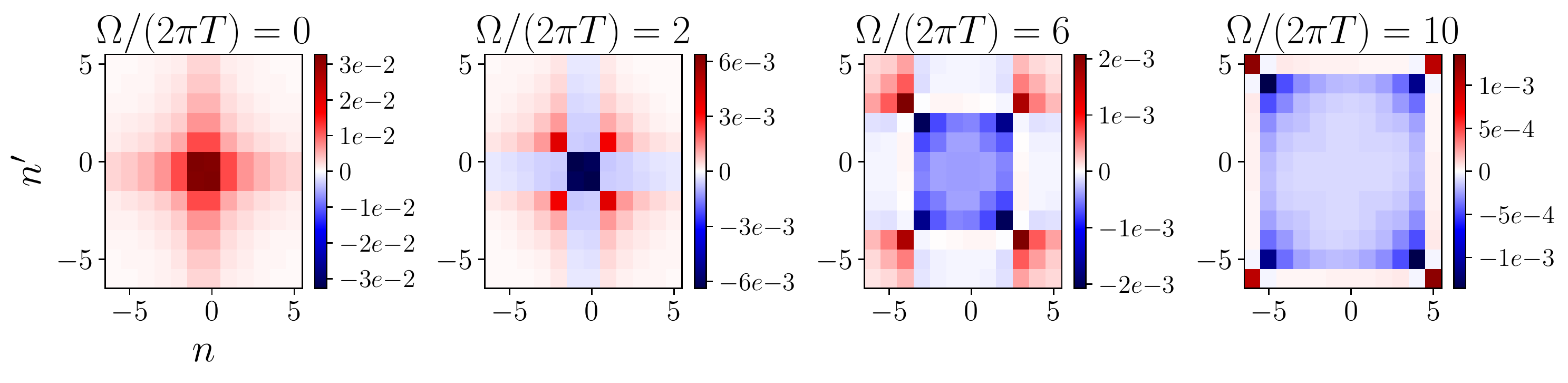}
    \caption{Same as Fig.~\ref{fig: dwave for factorization 1} with $U=3$ instead. Note the increase of one order of magnitude in the absolute values as compared to Fig.~\ref{fig: dwave for factorization 1}.}
    \label{fig: dwave for factorization 2}
\end{figure*}

We finally include also data for the rest function at finite doping. 
In Figs.~\ref{fig: dwave for factorization 1} and \ref{fig: dwave for factorization 2} we show the $d$-wave contribution to the rest function for $U=2$ and $U=3$ (and $t'=-0.2$) respectively. 
We observe a considerable increase with the interaction, despite the overall small values in both cases. This explains the small effect of the rest function observed in the ($d$-wave) superconducting susceptibility, see Figs.~\ref{fig: chi doping U2 bis} and \ref{fig: chi doping U3}.

\clearpage
\bibliography{main.bbl}

\end{document}